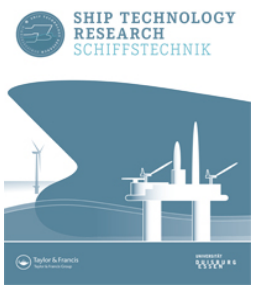

# Ship Technology Research

## Schiffstechnik

# An Artificial Bee Colony optimization-based approach for sizing and composition of Arctic offshore drilling support fleets considering cost-efficiency

A. A. Kondratenko, M. Bergström, M. Suominen & P. Kujala

**To cite this article:** A. A. Kondratenko, M. Bergström, M. Suominen & P. Kujala (2022): An Artificial Bee Colony optimization-based approach for sizing and composition of Arctic offshore drilling support fleets considering cost-efficiency, Ship Technology Research, DOI: 10.1080/09377255.2021.2022906

**To link to this article:** https://doi.org/10.1080/09377255.2021.2022906

Published online: 05 Jan 2022.

Submit your article to this journal 🗗

Article views: 1237

View related articles 🗗

View Crossmark data 🗗

Taylor & Francis
Taylor & Francis Group

# An Artificial Bee Colony optimization-based approach for sizing and composition of Arctic offshore drilling support fleets considering cost-efficiency

A. A. Kondratenko , M. Bergström , M. Suominen  and P. Kujala

School of Engineering, Department of Mechanical Engineering, Aalto University, Aalto, Finland

**ABSTRACT**
This article presents an optimization-based approach for sizing and composition of an Arctic offshore drilling support fleet considering cost-efficiency. The approach studies the main types of duties related to Arctic offshore drillings: supply, towing, anchor handling, standby, oil spill response, firefighting, and ice management. The approach considers the combined effect of the expected costs of accidental events, the versatility of individual support vessels, and ice management. The approach applies an Artificial Bee Colony algorithm-based optimization procedure. As demonstrated through case studies, the approach may help to find a range of cost-efficient fleet compositions. Some of the obtained solutions are similar to corresponding real-life fleets, indicating that the approach works in principle. Sensitivity analyses indicate that the consideration of the expected costs from accidental events significantly impacts the obtained solution, and that investments to reduce these costs may improve the overall cost-efficiency of an Arctic offshore drilling support fleet.

**Abbreviations:** ABC: Artificial bee colony; AH: Anchor handling; DP: Dynamic positioning; Fi-Fi: Firefighting; KPI: Key performance indicator; NM: Nautical mile; OSV: Offshore support vessel; POLARIS: Polar operational limit assessment risk indexing system; PC: Polar class; RIO: Risk index outcome; RS: Russian maritime register of shipping; WMO: World meteorological organization

**ARTICLE HISTORY**
Received 21 December 2021
Accepted 21 December 2021

## 1. Introduction

The Arctic contains very significant oil and gas reserves. As per (U.S. Geological Survey 2008), the known oil and gas fields in the Arctic account for some 240 billion barrels of oil and oil-equivalent natural gas, which is around 10% of the world's known conventional petroleum resources. Most of the Arctic, especially offshore, remains essentially unexplored with respect to fossil fuels. As per (U.S. Geological Survey 2008) and (EIA 2012), the unexplored Arctic reserves may account for an estimated 13% (90 billion barrels) of the world's undiscovered conventional oil resources and 30% of its undiscovered conventional natural gas resources.

Current Arctic offshore oil and gas exploitation activities are mainly limited to Norwegian and Russian waters. In the Norwegian Arctic, the main developments are the Snøhvit and Goliath fields in the Barents Sea. A third field– the Johan Castberg field – is currently being developed around 100 km northwest of the Snøhvit field. Production there is due to start in 2022 (Equinor 2018). As per (Reuters 2020), exploitation drilling aiming to develop additional new fields on the Norwegian Arctic shelf is expected in the future. In the Russian Arctic, presently, there is a single offshore field under development, known as the Prirazlomnoye field, which is located on the Pechora Sea shelf. The Prizazlomaya field is the only Arctic field in year-round operation using an ice-strengthened offshore platform (Tarovik et al. 2018). Additional fields are expected to be developed in the future as exploration drilling is taking place in the Kara Sea. In the American Arctic, on the other hand, there is currently no offshore drilling. However, this could change as there are strong economic incentives to develop the rich oil and gas fields in the area.

Offshore exploration drilling requires an offshore drilling rig, which is a large complex structure. For safe and efficient operations, a drilling rig requires continuous support from an offshore drilling support fleet. Typical drilling support fleet duties include supply, towing, anchor handling (AH), crew transportation, safety standby, as well as emergency functions such as firefighting (Fi-Fi) and oil recovery. A drilling support fleet must also deal with ice management in ice-infested waters, aiming to protect drilling rigs from dangerous ice interactions.

In the Arctic, the length of the drilling season is typically 3–5 months, occurring within the period July – November, when the presence of ice is minimal. A typical exploration drilling process can be chronologically divided into three stages: mobilization, operation, and demobilization. Support fleet duties related to mobilization and demobilization consist mainly of towing and AH, whereas the operation stage support mainly consists of supply. A drilling support fleet typically consists of several offshore support vessels (OSVs), including both multi- and single-purpose vessels.

Due to the complexity of the operations and the harsh but fragile Arctic environment, an offshore exploration drilling operation in the Arctic is typically subject to significant safety, environmental, and economic risk. Because the support fleet has a strong influence on the reliability and robustness of the operations, it has a key role in managing such risks. However, because support fleet-related costs are high, operators must find a balance between costs and risks.

Any offshore drilling installation has an upper limit regarding the ice conditions in which it can operate. Because drilling installations are unable to maneuver to avoid ice, ice management is paramount. This applies especially to the start and end of the drilling season when the probability of ice and iceberg occurrence is the highest. If there is a possibility that the limit will be exceeded, the installation must be evacuated and relocated from the drilling point in advance with the help of tugs and icebreakers, resulting in high financial losses. It is possible to extend the range of safe operating conditions by increasing the number of icebreakers to assist in the ice management around the installation (El Bakkay et al. 2014). However, the cost of ice management resources is also high. Thus, finding an appropriate ice management solution requires a balanced consideration of, among others, the possible ice scenarios, characteristics of the installation, and the consequences of an evacuation (Keinonen 2008).

In regions with significant offshore activity, such as the Norwegian Sea or the North Sea, a drilling support fleet is often organizationally divided into monofunctional sub-fleets that perform the same duty for multiple installations. As a result, such sub-fleets can be optimized for a specific type of duty. However, the distance between installations might be significant in remote regions with limited offshore activity, such as most Arctic regions. As a result, a drilling support fleet must perform multiple duties for a single installation (Gauthier and Molyneux 2018), (Kondratenko and Tarovik 2020). For such cases, the versatility of individual vessels is essential.

Existing methods for sizing and composition of offshore support fleets are mainly developed for regions with significant offshore activity and involve optimization of monofunctional sub-fleets. Early works include (Fagerholt and Lindstad 2000), which addresses monofunctional supply fleet sizing and composition, focusing on brute-force optimization of supply schedules. The study by (Fagerholt and Lindstad 2000) initiated a significant amount of further research: (Aas et al. 2007) considered storage capacity constraints, (Halvorsen-Weare et al. 2012) considered regular weekly schedules, and (Norlund et al. 2015) considered schedule robustness. Additional optimization criteria (e.g. travel time, OSV charter, and operation costs) were considered by (Aas et al. 2007), (Halvorsen-Weare et al. 2012). Different types of optimization algorithms (e.g. Tabu search heuristic, the Genetic algorithm) were adopted by (Gribkovskaia et al. 2008), (Borthen et al. 2018). Detailed operating conditions (e.g. wind, waves, and sea ice) were considered by (Maisiuk and Gribkovskaia 2014). Detailed models for studying local aspects of support fleet operations were presented by (Wu and Moan 2017), (Tarovik et al. 2018), (Lu et al. 2019), (Mazurek 2019). The versatility of an offshore support vessel from the ship design perspective was studied by (Rehn et al. 2018).

A significant limitation of the existing methods and approaches is that none of them is applicable to multifunctional (versatile) Arctic offshore drilling support fleets. The optimization of the sizing and composition of such a support fleet represents a complex mathematical problem, generally formulated as a constrained integer nonlinear programming problem. The challenge is further complicated by a large number of potential solutions. For instance, considering a realistic number of different candidate vessel types (about thirty or higher) for a typical Arctic offshore drilling support fleet, there might be more than $3 \cdot 10^{12}$ alternative fleet configurations.

In the optimization of multifunctional Arctic offshore drilling support fleet, like in most real-world optimization problems, the objective function is complex, non-linear, highly constrained (discontinuous), full of logical operators, and consequently cannot be differentiated. The optimization process is further complicated by the objective function's multiple local optima. As a result, the objective function must be operated as a black box by an optimization algorithm that can work with no information available about the target function. Classic optimization approaches such as the simplex or gradient descent method can not be applied for this optimization problem.

As demonstrated by (Fagerholt and Lindstad 2000), (Aas et al. 2007), (Halvorsen-Weare et al. 2012), some early attempts using straightforward optimization algorithms like brute force search, as well as commercial nonlinear problem solvers, for the optimization of offshore support fleets have required significant model simplifications to keep the calculation time reasonable.

As a result, the practical utility and value of the approaches appear limited. Many recent studies, e.g. (Gribkovskaia et al. 2008), (Borthen et al. 2018) and (Ehlers et al. 2019), demonstrated that various meta-heuristic optimization algorithms, such as the Tabu search heuristic, Genetic algorithm, and particle swarm algorithm, enable fast and reliable optimization of highly-constrained offshore support fleet sizing and composition problems. However, although it is known that metaheuristics are preferable for support fleet optimization, there are different opinions on which metaheuristic is the most efficient for this task.

One of these alternative methods is the Artificial Bee Colony algorithm (ABC) – a swarm-based metaheuristic whose behaviour is like honeybees. In comparison with other optimization metaheuristics, such as the genetic algorithm, differential evolution algorithm, and particle swarm algorithm, the ABC algorithm has demonstrated significant advantages in terms of performance and simplicity, especially when dealing with real-life optimization problems with high computational complexity and non-linear behaviour of the modelled objects (Karaboga and Akay 2009), (Akay 2013). Based on these promising findings, we adapted the ABC algorithm to the optimization problem at hand.

The development of new approaches for Arctic offshore drilling support fleet optimization, addressing the above-described limitation of existing models and approaches, is well motivated by the needs of the industry (RAO/CIS Offshore 2019) (Gauthier and Molyneux 2018). Following the current practice in fleet sizing and composition, only supply and crew transportation requirements and corresponding Arctic offshore drilling support fleet capabilities undergo a quantitative assessment, while other functions are evaluated based on requirements lumped together, without a specific level of detail, based on the subjective opinions of individual experts (Ehlers et al. 2019). It is noted that the experts have no consolidated opinion on these matters, and that their requirements are often contradictory (RAO/CIS Offshore 2019). Recent studies (Gauthier and Molyneux 2018) (Ehlers et al. 2019) presented approaches for Arctic offshore support fleet optimization based on the existing paradigm considering functionality requirements (towing, AH, safety standby, ice management) as predefined entities. These requirements are limited to simplified constraints: the fleet must have one or more vessels with a specific functionality.

To support the sustainable development of Arctic oil and gas, this article proposes an optimization-based approach for sizing and composition of cost-efficient drilling support fleets for offshore exploration in the Arctic. The proposed approach assesses a fleet's functionality quantitatively, considering current offshore marine practices. The quantitative assessment is based on a novel formulation of the support fleet optimization problem, considering the expected costs of accidental events, support vessel versatility (i.e. the ability of an individual vessel to perform various duties, including ice management), as well as current statistics and offshore regulations. The proposed approach aims to facilitate a systematic search for a support fleet configuration (number of vessels of different types) that minimizing the overall expected costs while meeting set criteria for safety and performance using a limited amount of data typical for Arctic offshore logistics management.

In this context, the study addresses the following research questions: Does a quantitative assessment of fleet functionality contribute to a systemic and integrated search for an optimal Arctic offshore drilling support fleet? Does the consideration of the expected costs of accidental events significantly impacts the choice of support fleet composition? Is it economically motivated to invest in safety? Is it feasible to apply an ABC algorithm to optimize an Arctic offshore drilling support fleet?

As indicated by the research questions, the study has an exploratory nature, with the aim to make a first step towards developing a holistic approach that reduces the level of subjectivity in high-level Arctic offshore support fleet sizing and composition. Related topics not addressed include stochastic simulation, ship-ice interaction and resistance modelling, and offshore crew transportation. Ship-ice interaction-related issues are not specifically addressed as Arctic offshore drilling operations typically are carried out in summer when the occurrence of sea ice is rare. Crew transport is not specifically addressed as the typical passenger capacity of an Arctic OSV of around 30 persons is high in relation to the typical crew size of an Arctic drilling installation of around 130 persons. The study does not include any holistic risk assessment and considers only those hazards that are dependent on the fleet configuration; other hazards, e.g. icing, grounding, loss of stability, are not specifically considered.

## 2. Theory and methods

### 2.1. Problem formulation

The objective function of the optimization process estimates the total costs of operations and is determined as per Equation (2.1). The objective function is minimized using the ABC algorithm (see Sec. 2.8). Candidate vessels are selected from the vessel pool $V$ that includes a specific number of vessels of each predefined vessel type.

Equation (2.1) contains different types of risk-associated costs, which are added to the normal charter costs ($C_{v,d}^{cr}$) and fuel costs ($C_{v,d}^{f}$). All risks $C^{k}$, where $k$ is the type of risk, are divided into three main categories: asset risks $C_{v,d}^{ra}$, human life risks $C_{v,d}^{rh}$, and risks specifically related to the presence of ice $C_{ib}^{ri}$. The objective function is calculated as the sum of the costs of the various duties performed by each vessel.

Equation 2.2 determines duty requirements for each vessel $v$ aiming to execute duty $d$. Parameters and constraints of Equation (2.1–2.5) are defined as per the methods described in Sec. (2.2–2.5).

The risk management procedure is handled at two different levels. At the first level, constraints set by Equation (2.3–2.5) are applied. These concern the minimum number of vessels $n_d^{min}$ and their characteristics that are needed to safely perform a duty $d$, as well as the minimum acceptable Risk Index Outcome (RIO) value (see Sec. 2.7) to assess limitations for operations in ice. As per Equation (2.5), as a conservative criterion, only RIO values indicating 'normal' operations e accepted. At the second level, the impact of various fleet configurations $i \in I$ on the risk values is measured ($I$ is the set of feasible fleet configurations). In general, the influence of any vessel $v$ on $i$ is nonlinear and therefore different for various $i$.

The study is limited to risks that are dependent on the fleet configuration. These risks are defined as $R_d^k = F_d^k \, E_d^k$, where $F_d^k = f(i)$ is the frequency of an event occurrence, and $E_d^k = f(i)$ is the corresponding consequence (economic loss) in USD. The individual risk contribution $C_{v,d}^k$ of an OSV (see Equation 2.1) is calculated as $R_d^k / \sum_{v \in V} x_{v,d}$. The same approach is applied for an icebreaker with the contribution $C_{ib}^{ri}$.

Risks associated with asset losses are expressed in repair cost, cost of new assets, or lost profit. The same applies to risks associated with the presence of ice, which are considered as lost profit and wasted resources due to unscheduled operational interruptions.

An economic-based approach is applied to assess the risk of human losses (Viscusi and Gayer 2002). As per this approach, a risk-reducing measure is motivated as long as its cost of averting a statistical fatality is below the value of a statistical life.

Sets:

$V$: Available OSVs, indexed by $v$

$IB$: Available icebreakers, indexed by $ib$

$D$: Required duties (supply, towing, anchor handling, safety standby, firefighting, oil recovery) to support the platform, indexed by $d$

$N_d^{min}$: Set of minimum numbers of vessels to perform duty $d$, indexed by $n_d^{min}$

$P_{v,d}$: Set of performance capacities to execute duty $d$ by vessel $v$, indexed by $p_{v,d}$

$R_d$: Set of functional specific requirements to the fleet for successful performing duty $d$, indexed by $r_d$

Parameters:

$C_{v,d}^{cr}$: Charter cost for vessel $v$ to perform duty $d$

$C_{v,d}^f$: Fuel cost for vessel $v$ to perform duty $d$

$C_{v,d}^{ra}$: Contribution of vessel $v$ performing duty $d$ to the risk of asset loss

$C_{v,d}^{rh}$: Contribution of vessel $v$ performing duty $d$ to the risk of loss of human life

$C_{ib}^{ri}$: Contribution of icebreaker vessel $ib$ to the risk of economic loss due to ice

$$\alpha_{v,d} = \left\{ \begin{array}{c} 1, \text{ if vessel } v \text{ is specified to perform duty } d \\ 0, \text{ otherwise} \end{array} \right\}$$

$RIO_v$: Risk Index Outcome to assess limitations for operations in ice

Variables:

$$x_{v,d} = \left\{ \begin{array}{ll} 1 & \textit{if vessel } v \textit{ performs duty } d \\ 0 & \textit{otherwise} \end{array} \right.$$

$$\min \sum_{v \in V} \sum_{d \in D} x_{v,d} (C_{v,d}^{cr} + C_{v,d}^f + C_{v,d}^{ra} + C_{v,d}^{rh})$$
$$+ \sum_{ib \in IB} C_{ib}^{ri}, \tag{2.1}$$

$$x_{v,d} \neq 1 \text{ if } \alpha_{v,d} \neq 1, v \in V, d \in D, \tag{2.2}$$

$$\sum_{v \in V} x_{v,d} \geq n_d^{min}, \ d \in D, \ n_d^{min} \in N_d^{min} \tag{2.3}$$

$$\sum_{v \in V} x_{v,d} p_{v,d} \geq r_d, \ v \in V, \ d \in D, \ p_{v,d}$$
$$\in P_{v,d}, \ r_d \in R_d, \tag{2.4}$$

$$RIO_v \geq 0, v \in V. \tag{2.5}$$

To improve computing efficiency, the optimized process is divided into two stages. In the first stage (see Sec. 2.4), optimization of the 'traditional' drilling support fleet, responsible for supply, towing, AH, safety standby, Fi-Fi, and oil recovery, is carried out. In the second stage (see Sec. 2.5), considering the outcome of the optimization of the 'traditional support fleet', optimization of the 'ice management fleet' is carried out. The 'ice management fleet' is an independent part of the Arctic offshore support fleet, which is organized to modify the ice to extend the range of safe operating ice conditions. An Arctic offshore support fleet is formed from the optimal vessels of the 'traditional' drilling support fleet and the 'ice management fleet'.

Anyhow, the optimization process could also be described as a single process as the 'separation' of the processes is solely added to simplify the related mathematical operation. The separation does not impact the optimization results as the optimization objective of both the 'traditional support fleet' and the 'ice management fleet' are the same, namely cost-efficiency, and because the consequences of an undesired event for the 'ice management fleet' are defined as a linear function of the expenses associated with the 'traditional support fleet', meaning that any improvement in the objective of the 'traditional support fleet' always results in an improvement of the cost-efficiency of the 'ice-management fleet'.

## 2.2. Constraints

Equation (2.3–2.4) defines seven constraints concerning the minimum number of vessels required per duty and their required characteristics. These are listed in Table 1 and described in the following.

Constraint 1 requires the monthly supply capacity of the fleet $S_{rate}$ (the total available deck area, m² per month) to be equal or higher than the monthly consumption rate $Cons_{rate}$ of the installation. The value of $S_{rate}$ is determined considering redundancy (see Sec. 2.4). Constraint 2 determines the minimum machinery power $N_{pp, towing}^{min}$ (kW) of vessels involved in towing operations. The minimum required power is calculated as a function of the number of vessels involved in the task, towing speed, characteristics of the installation, and environmental conditions as per (Russian Maritime Register of Shipping 2014). The aim is to prevent a loss of control of an installation due to the lack of power – something that could have severe consequences (Necci et al. 2019). Constraint 3 determines that AH operation requires at least two vessels- one vessel to secure the anchor, and another vessel to keep the installation in position (Gibson 1999). The machinery power of the main anchor handling vessel must be at least $N_{pp, AH}^{min}$ (kW), which is determined as per (Gibson 1999). Meeting this constraint is important in terms of the safety of AH operation (Gibson 1999), (Laverick 2011). For secondary AH vessels, there is no specific criterion for $N_{pp}$, as any vessel certified to perform AH is assumed to have sufficient machinery power. Constraint 4 requires the continuous presence of at least one safety standby vessel in the vicinity of an installation as per existing industry standards (Coles et al. 2000), (Gauthier and Molyneux 2018). In the case of an evacuation, the capacity of a single certified safety standby vessel is assumed to be sufficient to provide shelter for the whole crew. Constraint 5 requires the fleet to include at least two vessels with Fi-Fi capabilities. Two vessels are needed both for redundancy, and to be able and to fight a fire on an installation simultaneously from two different directions. Constraint 6 requires the fleet to include at least one vessel with certified oil recovery capability, which is usually sufficient to manage the consequences of an oil spill at an early stage. Constraint 7 requires the presence of at least one vessel with an ice class higher than the minimum required by Equation (2.5). This is necessary for the safe demobilization of an offshore installation from its location in case of an emergency (See sec. 2.5, passive ice management scenario).

## 2.3. Categories of risk events

In this study, risks associated with an Arctic offshore drilling support activity are analyzed based on empirical data by (Oil & Gas UK 2009). This database represents offshore operations in harsh environments and includes data from multiple sources, including the Offshore Safety Division of the Health and Safety Executive (United Kingdom), SINTEF (Norway), DNV GL (Norway, Germany), and the Marine Accident Investigation Branch (United Kingdom). The applied database (Oil & Gas UK 2009) was formed by processing the raw accident data, revealing the chains of events that have resulted in an accident and the frequencies of event occurrences.

Categories of risk events and drilling support fleet parameters influencing those risk events were determined by analyzing the database as per Table 2. The risk event categories were determined as per (DNV GL 2013). Description of Fi-Fi and DP classes is provided in Table 3. The contents of Table 2 are explained in the following.

Different types of towing-related risk- events may occur when moving an installation from one location to another. The potential consequences are severe. For instance, a loss of control of a platform during a towing operation could result in the grounding or sinking of a rig (Necci et al. 2019). Such risks can be managed using multiple tugs for high redundancy. Thus, in this study, the risk of such accidents is assumed dependent on the number of tugs.

**Table 1.** Constraints applied in the optimization model.

| Nr. | Duty $d$ | $n_d^{min}$, see Equation (2.3) | $r_d$, see Equation (2.4) |
|---|---|---|---|
| 1 | Supply | 1 | $S_{rate} \geq Cons_{rate}$ |
| 2 | Towing | 1 | $N_{pp} \geq N_{pp, towing}^{min}$ $$N_{pp, towing}^{min} = f\left(\sum_{v \in V} X_{v, towing}\right)$$ |
| 3 | AH | 2 | At least one vessel with $N_{pp} \geq N_{pp,AH}^{min}$ |
| 4 | Safety Standby | 1 | Full time |
| 5 | Fi-Fi | 2 | – |
| 6 | Oil recovery | 1 | – |
| 7 | Ice management (if applicable) | 1 | At least one vessel with ice class = min ice class + 1 |

**Table 2.** Risk events, the corresponding support fleet duties, and the influencing parameters.

| Duty | Risk event (WOAD) | Fleet parameters |
|---|---|---|
| Towing | Towing/towline (Loss of control of platform) | Number of tugs |
| AH | Miscellaneous | Constraints compliance |
| Supply | Contact | DP class |
| Fi-Fi | Fire | Fi-Fi class |
| Safety standby | Miscellaneous | Constraints compliance |
| Oil recovery | Spill/release | Constraints compliance |

**Table 3.** Description of the firefighting and dynamic positioning classes (GL 2008) (IMO 1994).

| Class | 0 | 1 | 2 | 3 |
|---|---|---|---|---|
| Fi-Fi | – | Fighting only the initial stage of fire in the vicinity of the installation. | Sustained fighting of large fires and cooling parts of the installation. | Class 2 is complemented by greater fire extinguishing capacity and additional equipment. |
| DP | – | No redundancy. Loss of position due to a single fault. | No single fault in an active system results in position loss. | No position loss during and following any single fault, including loss of a compartment. |

Supply operations are associated with a high risk of collision as they require proximity between the involved vessel and offshore installation. In this study, following (IMCA 2015), a ship's DP class is assumed to affect both the frequency and consequences of unintended contact events. Another factor affecting the risk of a collision is the frequency at which supply vessels visit an installation $n_{spw}$ (Aas, Halskau, & Wallace, The role of supply vessels in offshore logistics, 2009). The frequency of visits ($n_{spw}$) depends both on the rate of consumption ($Cons_{rate}$) and the capacity of an installation $A_I$ (m²).

A fire onboard an installation can rapidly escalate into a catastrophic event (Halim et al. 2018). The main objective of a support vessel certified for Fi-Fi activity is to fight a fire at an early stage or to suppress a fire until a specialized firefighting fleet has arrived. In this study, as per Table 3, a vessel's firefighting capability is quantified in terms of its Fi-Fi class.

The resources for other duties, including AH, safety standby, and oil recovery, are assumed fixed. For these vital duties, the minimum criteria as per Table 1 must be met.

## 2.4. Evaluation of the traditional support fleet performance

The traditional support fleet performs duties 1–6 (see Table 1). The consequences of risk events associated with towing and supply duties (Table 4) are categorized and assessed following (DNV GL 2013) (GL 2008). Accordingly, economic losses are determined in terms of loss equivalents, estimated considering the actual cost of repair works, delivery of spares, and operational downtime for an installation with an assumed price of USD 700 million (Kaiser and Snyder 2012). As per (Health and Safety Executive 2019), accidents involving offshore support vessels typically have a limited impact on the environment, resulting in minor clean-ups after potential collisions (included in the economic losses estimation). Major significant oil spills in the offshore practice are related to the individual operation of an offshore installation or large transport vessels, which is outside the study's scope. Thus, the environmental impact of offshore support fleet operation is assumed to be mainly associated with exhaust gas emissions related to regular fuel consumption.

**Table 4.** Damage categorization with approximate losses equivalents.

| Damage | Insignificant | Minor | Severe |
|---|---|---|---|
| Description | Damage to towline, thrusters, generators, and drivers | Minor damage to single essential equipment | Severe damage to one or more modules of the unit |
| Assets losses, USD | 200,000 | 1 million | 70 million |
| Human losses (fatalities) | 0 | 0.2 | 2 |

Human losses are estimated conservatively based on (Gunter et al. 2013), (Ibrion et al. 2020) using the Equivalent Fatality Concept (IMO 2015). Accordingly, for instance, two human fatalities equal 20 cases of severe injury or 200 cases of minor injuries. Because the uncertainty in the assessment of risk consequences is substantial, in Section 3.5 we present a sensitivity analysis of the related parameters.

Towing-related costs, including accident costs, are calculated as per Equation (2.6–2.9); $C^{cr}_{v,towing}$ is the charter cost for vessel v to perform towing; $C^f_{v,towing}$ is the fuel cost for vessel v to perform towing.

$$C^{cr}_{v,towing} = CR_v t_{tow},\qquad(2.6)$$

where $CR_v$ is the daily charter rate in USD (see Sec. 2.6), $t_{tow} = 2dist_{tow}/(24v_{tow})$ is the total towing time in days, considering two transfer voyages (shore – drilling point -shore); $dist_{tow}$ is the towing distance from the port to the drilling location in nautical miles; $v_{tow}$ is the towing speed in knots.

$$C^f_{v,towing} = p_{fuel}(24t_{tow}k_{red}N^{min}_{pp,towing}q),\qquad(2.7)$$

where $p_{fuel}$ is the fuel price in USD, $k_{red} = 0.9$ is the reduction coefficient, which considers the difference between the actual power used in operation and the nominal required power $N^{min}_{pp,\,towing}$; $q = 0.221\cdot10^{-3}$, t/ kW·h is the specific fuel consumption.

Towing risk losses associated with assets $R^{ra}_{towing}$ and human lives $R^{rh}_{towing}$ are calculated for the whole fleet as per Equation (2.8–2.9).

$$R^{ra}_{towing} = \frac{F_{towing}E^{ra}_{towing}(t_{op} + t_{tow} + t_{AH})}{365},\qquad(2.8)$$

$$R^{rh}_{towing} = \frac{F_{towing}E^{rh}_{towing}(t_{op} + t_{tow} + t_{AH})}{365},\qquad(2.9)$$

where $F_{towing} = 0.03$ is the towing/towline risk event occurrence frequency per unit per year; $E^{ra}_{towing}$ is the assumed value of assets losses related to a towing/towline risk event. The severity of the consequences (see Table 4) is assumed to be dependent on the number of tugs involved in the operation as follows (DNV GL 2013) (Necci et al. 2019):

- Scenario A: Insignificant damage in the case of three or more tugs, ensuring towing force in two directions following a fault disabling one of the tugs. Consequences are assumed limited to those caused directly by the occurred fault.
- Scenario B: Minor damage in the case of two tugs (sufficient redundancy to prevent the complete loss of control of an installation, but a fault may nevertheless result in reduced control of an installation due to a single direction of towing force).
- Scenario C: Severe damage in the case of a single tug (no redundancy, any fault disabling the single tug will result in a loss of control of the installation).

The sum of the time spent to support the drilling operations ($t_{op}$), towing operations ($t_{tow}$), and anchor handling operations ($t_{AH}$) represents the total active time of an installation during the considered exploration project (one season). In this study, the values of $t_{op}$ and $t_{AH}$ are treated as inputs. $E^{rh}_{towing}$ is the human life loss, determined as the product of the number of human fatalities (see Table 4) and the assumed value of a human life.

AH related costs are calculated as per Equation (2.10–2.11), based on the charter price $C^{cr}_{v,d}$ and fuel cost $C^{f}_{v,d}$ of each AH vessel. The anchor handling time ($t_{AH}$) represents the assumed duration of two AH operations.

$$C^{cr}_{v,AH} = CR_v t_{AH}, \qquad (2.10)$$

$$C^{f}_{v,AH} = p_{fuel} t_{AH} \qquad (2.11)$$

where $c_{AH}$ is the fuel consumption rate [ton/day] during an AH operation.

Costs related to supply duties are calculated as per Equation (2.12–2.13), based on the charter price $C^{cr}_{v,d}$ and fuel cost $C^{f}_{v,d}$ of each supply vessel.

$$C^{cr}_{v,supply} = CR_v t_{op}, \qquad (2.12)$$

$$C^{f}_{v,supply} = \frac{p_{fuel} t_{op}}{t_{v,voyage} c_{v,voyage}}, \qquad (2.13)$$

where $t_{v,\,voyage} = t_{v,p} + t_{v,pl} + 2t_{v,mov}$ is the duration of a return voyage for a vessel $v$ in days; $c_{v,\,voyage}$ is fuel consumption [ton] of a vessel $v$ per voyage; $t_{v,p} = 1.426 + 0.0005DW_v$ is the duration of loading and unloading operations at the supply base (Kondratenko and Tarovik 2020); $t_{v,pl} = 0.6 + 0.00021DW_v$ is the duration of loading and unloading at the installation

(Kondratenko and Tarovik 2020); $t_{v,mov} = dist_{sup} /(24v_{cr})$; $dist_{sup}$ is the distance from the supply base to the drilling point in nautical miles; $v_{cr}$ is the average cruising speed [knots] of the vessel during a voyage. The cruising speed is an important parameter as it impacts the overall supply fleet performance and should be calculated considering wind and wave characteristics. For this, we apply an approach proposed by (Kwon 2008), which estimates the loss in vessel speed due to wind as a function of $f(BN, C_b, \Delta$ $and\ Fr)$, where BN is the Beaufort number, $C_b$ is the vessel's block coefficient, $\Delta$ is the vessel's displacement in tons, $Fr$ is the Froude number.

The total fuel consumption of a vessel per voyage is calculated as per Equation (2.14) as the sum of fuel consumption for cargo operations and moving between the supply port and the installation.

$$c_{v,voyage} = t_{v,p} c_{v,p} + t_{v,pl} c_{v,pl} + 2t_{v,mov} c_{v,mov}, \qquad (2.14)$$

where $c_{v,p}$ is the fuel consumption rate [tons/day] for cargo operations at the supply port; $c_{v,pl}$ is the fuel consumption rate [tons/day] for cargo operations at the installation. The values of $c_{v,p}$ and $c_{v,pl}$ could be taken from the vessel specification and do not differ significantly between vessels. Parameter $c_{v,mov}$ represents the vessel's fuel consumption at cruising speed. According to (MarineTraffic 2020), the maximum cruising speed of supply vessels rarely exceeds 10 knots, which is set as the default value of $v_{cr}$ for operations when $BN = 0$ and applied for the estimation of $c_{v,mov}$. The fuel consumption $c_{v,mov}$ varies between different vessels, and the variation is particularly significant between open water and ice-going vessels. In this study, the fuel consumption rate of different vessels was calculated based on their technical characteristics (see Table A2) as per a methodology by (Kondratenko and Tarovik 2020).

The transport capacity of the support fleet is expressed by the supply rate $S_{rate}$, which is determined as per Equation (2.15) in $m^2$ per month.

$$S_{rate} = \sum_{v \in V} x_{v,supply} S_{rate,v}, \qquad (2.15)$$

where $x_{v,supply}$ is a Boolean variable that equals 0 if a vessel does not perform supplies and 1 otherwise; $S_{rate,\,v}$ is the supply rate of vessel $v$ in $m^2$ per month, calculated by Equation (2.16).

$$S_{rate,v} = \frac{30S_{v,useful}}{t_{v,voyage}} \qquad (2.16)$$

For vessels engaged in a supply activity, the deck area used for supplies $S_{v,\,useful}$ is determined as per Equation (2.17).

$$S_{v,useful} = \left\{ \begin{array}{l} S_v k_s, \ if \ S_v k_s \leq S_{deck} \\ S_{deck}, \ if \ S_v k_s > S_{deck} \end{array} \right\}, \qquad (2.17)$$

where $S_v$ represents the nominal useful deck area of a vessel $v$; $k_s = 0.7$ is a safety factor considering the partial utilization of a cargo deck area (15%) and redundancy (15%). $S_{deck}$ represents the cargo deck area of the installation. Following Equation (2.17), there must be enough storage area on the deck of an installation for all cargoes supplied by a vessel. This is an important constraint as exploration drilling rigs often have a limited deck area, which may decrease the utilization of large supply vessels. To obtain high operational reliability, which is vital in offshore drilling operations, for the calculation of the voyage time $t_{v,voyage}$ we used conservative assumptions (90 percentiles) for the wind speed and wave height (Liu et al. 2016).

Relationships between supply vessel DP class, the basic frequency of collisions between a supply vessel and an installation $F_{supply, 0}$, and the related consequences are determined as per Table 5 based on:

- Accident database (Oil & Gas UK 2009);
- Data provided in Table 3;
- Statistics and descriptions of individual collisions (Health and Safety Executive 2019).

As per Table 5, vessels of DP class 3 are assumed to be able to maintain a safe distance to an installation in all possible scenarios, as such vessels are assumed to have superior maneuverability in comparison with vessels of DP class 0–2, and the ability to maintain their position following a technical fault in any of their systems, resulting in negligible collision risk. Vessels of DP class 0–2 are assumed to have an equal level of maneuverability and hence the same collision frequency, but the consequences of a potential collision are assumed different. Most modern OSVs have DP class 2 and can hold their position in some cases (see Table 3) (IMO 1994), which according to the accident statistics, typically is enough to limit the consequences to insignificant damage, but minor damage is nevertheless possible. For vessels of DP class 1–2, accidents resulting in minor damage are assumed to be associated with a loss of position, which in the case of vessels of DP class 1 occurs following a single fault (see Table 3). Vessels of DP class 0 are operated manually without any automatic positioning system. Collisions involving such vessels are likely to have severe consequences due to human factors.

The frequency of collisions between a supply vessel and an installation $F_{supply}$ is calculated as per Equation (2.18), considering the number of visits to an installation per week.

$$F_{supply} = \frac{F_{supply, 0} n_{spw}}{n_{spw,0}}, \qquad (2.18)$$

where $n_{spw} = 7 Cons_{rate}/(30 S_{useful})$ is the average number of supply vessel visits to an installation per week (seven days); $S_{useful}$ is the average value of $S_{v,useful}$ for all supply vessels. In this study, the average weekly number of supply vessel visits to an installation ($n_{spw,0}$) is assumed to be 2 as per (Aas et al. 2009).

The risk of asset losses related to supply operations is calculated as per Equation (2.19).

$$R_{supply}^{ra} = \frac{\sum_{i=0}^{3} R_{supply,DPi}^{ra} n_{DPi}}{\sum_{v \in V} x_{v,supply}}, \qquad (2.19)$$

where $R_{supply, DPi}^{ra}$ represents the accident consequences of vessels of DP class $i$, and $n_{DPi}$ represents the number of supply vessels of DP class $i$. $R_{supply, DPi}^{ra}$ value is determined as per Equation (2.20).

$$R_{supply,DPi}^{ra} = \frac{\sum_{j \in J} F_{supply,DPi,j} E_j^{ra} (t_{op} + t_{tow} + t_{AH})}{365}, \qquad (2.20)$$

where $J$ is a set of consequence types $j$ (insignificant, minor, severe), $F_{supply, DPi, j}$ is the frequency of a supply risk events with consequences type $j$, calculated for a vessel with DP class $i$; $E_j^{ra}$ is the value of asset losses associated with consequences type $j$. The value of $R_{supply}^{rh}$ is calculated in the same manner as $R_{supply}^{ra}$, but by replacing $E_j^{ra}$ with $E_j^{rh}$, representing human losses.

The consequences of a fire are assumed to depend on the available firefighting resources as per Table 6 (Halim et al. 2018), (Paik et al. 2011). Specifically, the economic loss associated with a fire $R_{Fi-Fi}^{ra}$ is calculated as per Equation (2.21).

$$R_{Fi-Fi}^{ra} = \frac{F_{fire} E_{fire}^{ra} (t_{op} + t_{tow} + t_{AH})}{365}, \qquad (2.21)$$

where $F_{fire}$ is the annual occurrence frequency of a fire on an installation; and $E_{fire}^{ra}$ is the assets losses associated with fire as determined by Table 6. The corresponding risk to a human life $R_{Fi-Fi}^{rh}$, is calculated as per Equation (2.21), but by replacing $E_{fire}^{ra}$ with $E_{fire}^{rh}$, which is the product of the number of fatalities (see Table 6) and the assumed value of human life.

The final step to evaluate the support fleet performance is to estimate the charter price and fuel price for safety standby vessels as per Equation (2.22–2.23) for the total operational time.

**Table 5.** Basic frequencies [events per year] of vessel-installation collisions for various vessel DP classes and damage categories.

| Damage | Insignificant | Minor | Severe |
|--------|---------------|-------|--------|
| DP 0 | 0 | 0 | 0.217 |
| DP 1 | 0 | 0.217 | 0 |
| DP 2 | 0.195 | 0.022 | 0 |
| DP 3 | 0 | 0 | 0 |

**Table 6.** Different fire accident scenarios and approximate loss equivalents.

| Scenario | Scenario A: Insignificant loss (two or more vessels with Fi-Fi class 3) | Scenario B: Average loss (two or more vessels with Fi-Fi class 2) | Scenario C: Severe loss (two or more vessels with Fi-Fi class 1) |
|---|---|---|---|
| Description | Fire is suppressed in the initial stage | Prevention of the fire escalation until the arrival of specialized firefighting vessels | Rapid fire escalation resulting in substantial damage to the installation |
| Assets losses (USD) | 100000 | 142000 | 7000000 |
| Human losses | 0.13 | 0.17 | 8.9 |

$$C^{cr}_{v,SS} = CR_v(t_{op} + t_{tow} + t_{AH}), \quad (2.22)$$

$$C^{f}_{v,SS} = p_{fuel}(t_{op} + t_{tow} + t_{AH})c_{SS,} \quad (2.23)$$

where $c_{SS}$ is the fuel consumption rate [tons/day] for a safety standby operation.

## 2.5. Evaluation of the ice management fleet performance

The most severe ice conditions in which an installation can operate is quantified in terms of equivalent ice thickness $h_{eq}$ as per Equation 2.24. (CNIIMF 2014).

$$h_{eq} = c(h_i + 0.25bh_i + k_{sn}h_{sn}), \quad (2.24)$$

where $c$ is the ice concentration; $h_i$ is the level ice thickness [m]; $b$ is the amount of ice ridging on the so-called ball-scale; $k_{sn} = 0.5$ if $h_{sn} \geq 0.5$ m and $k_{sn} = 0.33$ otherwise; $h_{sn}$ is the snow cover thickness [m].

To determine a link between the configuration of the ice management fleet and its ability to reduce the value of $h_{eq}$, we apply a strategy-based approach as per Table 7. Specifically, three different strategies are considered: (A) complete ice management, (B) active ice management, and (C) passive ice management. In strategy C, the installation leaves the drilling location in advance if ice forecasts indicate that a specific upper limit value ($h_{max}$) of $h_{eq}$ will be exceeded. In that case, a single vessel with a high ice class (see Table 1, constraint 7) provides escort to a safe location. In strategy B, active ice management is provided by two vessels. One of the icebreakers has a higher ice class than required for independent operations in the prevailing ice conditions (see Table 1, constraint 7) and acts as the leader for the other vessel. The effectiveness of this strategy is assessed as per

(Keinonen 2008) in terms of the obtained equivalent reduction in ice thickness $h_{red}$. Considering a coefficient of determination $R^2$ of 0.95, we approximated this approach as per Equation (2.25).

$$h_{red} = \min\{0.0204EXP(1.9304h_{eq}); h_{eq}\}. \quad (2.25)$$

In strategy A, protection against all possible ice conditions is provided throughout the drilling period. This is achieved by a lead icebreaker, supported by four secondary vessels (Keinonen 2008).

The charter and fuel costs for the ice management vessels are estimated as per Equation (2.26–2.27).

$$C^{cr}_{v,IM} = CR_v(t_{op} + t_{tow} + t_{AH}), \quad (2.26)$$

$$C^{f}_{v,IM} = p_{fuel}(t_{op} + t_{tow} + t_{AH})c_{IM}, \quad (2.27)$$

where $c_{IM}$ equals the fuel consumption rate [tons/day] during the ice management activity.

Risks associated with the presence of sea ice are determined as per Equation (2.28).

$$R^{ri} = F^{ri}E^{ri}, \quad (2.28)$$

where $F^{ri}$ is the probability of an operational interruption due to sea ice and $E^{ri}$ is the financial consequences of an interruption.

We assume that operations might be interrupted by two different types of ice: difficult sea ice condition, and icebergs. Specifically, the probability of an operational interruption due to ice ($F^{ri}$) is estimated as per Equation (2.29).

$$F^{ri} = F^{ri}_0 + (1 - F^{ri}_0)p_{iceberg}, \quad (2.29)$$

where $F^{ri}_0$ is associated with sea ice, and the rest of the equation is associated with icebergs. The parameter $p_{iceberg}$ is the probability of iceberg occurrence during the drilling season. For ice management strategy A, the value of $F^{ri}$ is assumed to be 0. Parameter $F^{ri}_0$ is determined as per Equation (2.30).

$$F^{ri}_0 = \sum_{ic \in IC} i_{ic}F_{ic}, \quad (2.30)$$

where $IC$ is a set of possible ice condition types $ic$ (mild, average, severe); $F_{ic}$ is the assumed occurrence probability of ice condition of type $ic$; $i_{ic}$ is a Boolean value determined as per Equation (2.31), indicating whether a drilling interruption is needed in the ice

**Table 7.** Considered ice management strategies.

| Ice management strategy | Scenario A: Complete ice management | Scenario B: Active ice management | Scenario C: Passive ice management |
|---|---|---|---|
| Description | Protection of the drilling rig from impacts with ice and icebergs in all possible scenarios | Measures to reduce the $h_{eq}$ value | No active measures to affect the ice condition |
| Fleet | 5 vessels | 2 vessels | 1 vessel |

conditions $ic$.

$$i_{ic} = \begin{cases} 0, & if \ h_{red,ic} \leq h_{max} \\ 1, & if \ h_{red,ic} > h_{max} \end{cases}, \qquad (2.31)$$

where $h_{red,ic}$ is reduced ice thickness, which for the active ice management strategy is calculated following Equation (2.25). For the passive ice management strategy, $h_{red,ic}$ is assumed equal to $h_{eq}$.

It should be noted that the estimation of the financial consequences of a drilling interruption $E^{ri}$ is case-specific and thus difficult to assess. Anyhow, in this study, the value of $E^{ri}$ is assumed to correspond to the sum of the traditional fleet associated expenses from Equation (2.1) and the operational cost of the drilling installation calculated as per Equation (2.32).

$$E^{op}_{inst} = R^{op}(t_{op} + t_{tow} + t_{AH}), \qquad (2.32)$$

where $R^{op}$ is the daily charter rate of the installation, including operating expenses.

## 2.6. Time charter rate

To determine vessel charter rates (time charter), we use a methodology based on (Døsen and Langeland 2015). Accordingly, the daily charter rate $CR_v$ is determined in USD as per Equation (2.33). To account for that a vessel's icebreaking capabilities might significantly affect its charter rate, we introduced an additional coefficient $k_{IC}$ that we determine based on empirical data (Deynego and Gehaev 2019).

$$Ln(CR_v)$$
$$= k_{IC,v} \begin{pmatrix} \beta_0 + \beta_1 S_v + \beta_2 N_{pp,v} + \beta_3 DW_v + \beta_4 k_{DP2,v} + \beta_5 Age_v + \beta_6 Dur + \\ \beta_7 DF + \beta_8 K_p + \beta_9 K_d + \beta_{10} K_{br} + \beta_{11} P_{oil} + \beta_{12} P_{spot} + \beta_{13} O_{prod} \end{pmatrix},$$
$$(2.33)$$

where $k_{IC,v} = 0.55 \ h_{ice,v}+1$; $\beta_i$ ($i = 0..13$); $h_{ice,v}$, $S_v$, $N_{pp,v}$, and $DW_v$ are characteristics of vessel $v$; $k_{DP2,v}$ is a Boolean expression, which has a value of 1 if vessel DP class $\geq 2$, otherwise it has a value of 1 if $Age_v$ indicates a vessel's age, counted from its year launched; $Dur$ is the duration of a charter contract in days; $DF$ is the number of days from the contract initiation to the start date of the charter period; $K_p$ is a Boolean expression, which has a value of 1 if an assignment is limited to the production support, otherwise it has a value of 0 (this is a discount factor considering the level of difficulty of the operation); $K_d$ is a Boolean coefficient, which has a value of 1 if an assignment is limited to typical drilling support (this is another discount factor considering the level of difficulty of the operation); $K_{br}$ is the so-called Brazil coefficient, which has a Boolean value of 1 for operations in a developing market, and a Boolean value 0 for operations in a developed market. Lower charter rates in developed markets are associated with a high organization level and market oversaturation. $P_{oil}$

corresponds to the price per barrel of Brent crude oil; $P_{spot}$ is the monthly average spot charter rate; $O_{prod}$ is the worldwide monthly oil production.

## 2.7. Limitations for operation in ice

The ability of an OSVs to operate in ice is determined as per the Polar Operational Limit Assessment Risk Indexing System (POLARIS) (IMO 2016). Accordingly, whether a ship can safely operate in a specific area is determined based on its risk index outcome (RIO) value determined as per Equation (2.34) considering the vessel polar class (PC) (IMO 2016) and the prevailing ice conditions.

$$RIO_v = \sum_{i=1}^{12} C_i RIV_{v,i}, \qquad (2.34)$$

where $C_i$ is the concentration (in tenths) of an ice type $i$ within the ice regime; $RIV_{v,i}$ is the corresponding risk index value (RIV) for a vessel $v$ and an ice type $i$. Ice types conform to the World Meteorological Organization (WMO) nomenclature (World Meteorological Organization 2014). For ships of PC 1-7, a positive RIO value (RIO $\geq 0$) indicates 'normal' operations. A negative RIO value indicates 'elevated operational risk' if ($-10 \leq$ RIO $< 0$), or that the operation is 'subject to special consideration' if (RIO $< -10$). In this study, RIO value is calculated based on the worst possible ice conditions.

## 2.8. The ABC algorithm

In this study, the optimization of the fleet composition is performed by solving a nonlinear programming problem in the integer form, using a stochastic ABC algorithm based on (Karaboga and Basturk 2007). The behaviour of the ABC algorithm is modelled based on the foraging behaviour of a honeybee colony (Tereshko and Loengarov 2005). Accordingly, the model includes the following components:

(1) A passive set of potential food sources.
(2) A group of employed bees. Each employed bee flies to an arbitrary food source and sends information about its quality in terms of the distance from the hive, amount of energy, taste, and the level of energy needed to obtain the food.
(3) A group of temporary unemployed onlooker bees that aims to find a new food source based on the information provided by the employed bees.
(4) A group of scout bees that can independently find completely new food sources from the passive set of potential food sources.

If an onlooker bee, or a scout bee, finds a new food source that is better than any of the food sources occupying the employed bees, the new food source will replace the older one. Because the ABC algorithm in this manner integrates both local and global search methods, an efficient optimization process is obtained.

At the start of the optimizing procedure, the algorithm requires both initial input and control parameters. The initial input parameters describe the goal function $f(x)$, which assesses the quality of the solution and the dimension of the problem $D$. Here, the dimension of the problem consists of a set of variables $x_j$ ($j = 1..D$) describing the solution $i$ ($i = 1..N_{FS}$). In the present study, the dimension of the problem equals the considered number of different types of vessels. The control parameters consist of the colony size $C_S$ and the criterion of the calculation stop. The value of $C_S$ determines the number of food sources under consideration ($N_{FS}$) as determined by Equation (2.35). The number of bees of each category (employed, unemployed, and scouts) equals the $N_{FS}$ value. The criterion of the calculation stop is determined in an arbitrary form, for example, in terms of the maximum number of cycles, maximum calculation time, or in terms of a specific goal function value. The value of another important parameter, the limit value of scout $L$ as determined by Equation (2.36), is the function of $N_{FS}$ and $D$. This limit value determines the maximum number of attempts $n_{tr}$ to find a better food source based on the available food source data. To achieve an efficient optimization process, we also limit the maximum number of vessels of a specific vessel type.

$$N_{FS} = 0.5C_S. \qquad (2.35)$$

$$L = N_{FS}D. \qquad (2.36)$$

The complete optimization process can be summarized as follows (Karaboga and Akay 2009):

(1) Initialize the original food sources, each of which represents a fleet configuration.
(2) Place the employed bees on the food sources.
(3) Send the onlooker bees to the most promising food sources.
(4) Send the scouts to discover new food sources.
(5) Memorize the best-found food source.
(6) Repetition of steps 2–5 until the requirements are met.

Initialization of the original food sources (**operation 1**) is performed by creating a random array of size $N_{FS}{}^{\times}D$. In this study, each solution (food source) consists of $D$ number of non-negative integer values, which represent the number of vessels of each type in the current fleet configuration. Following the initialization, each food source's quality is evaluated using both a goal function and a fitness function. The goal function provides a value of the food source quality

$f_i$, and the fitness function transforms this value to the most appropriate form for the optimizing process. In the case of minimization, the fitness function $fit_i$ is as per Equation (2.37).

$$fit_i = \left\{ \begin{array}{l} \dfrac{1}{1 + f_i}, \, if \, f_i \geq 0 \\ 1 + abs(f_i), \, if \, f_i < 0 \end{array} \right\}. \qquad (2.37)$$

Once the initial food sources have been evaluated, the algorithm remembers the best solution and starts the optimization loop (**operation 2**). At this stage, each employed bee produces a new solution by randomly selecting the variable $x_{bj}$ (the number of vessels of type $j$ in a solution $i$) from its food source (i.e. the fleet solution) as described by Equation (2.38). In this study, Equation (2.38) was modified from its original form to produce a non-negative integer value.

$$x_{i,j}{}^{new} = |x_{i,j} + Round(F_{i,j}(x_{i,j} - x_{k,j}))| \qquad (2.38)$$

where $x_{i,j}{}^{new}$ is a new non-negative integer value, $Round$ is an operator that rounds a value to the nearest integer value, $F_{i,j}$ is a random number in the range $[-1, 1]$, and $k$ is a random index.

If the fitness value of the modified solution $i^{new}$ is higher than the initial one, $x_{bj}$ is replaced by $x_{i,j}{}^{new}$, and $n_{tr}$ gets a value of zero. In the opposite case, $x_{bj}$ is not modified, and the value of $n_{tr}$ is increased by one.

In the next stage (**operation 3**), each onlooker bee aims to improve one of the available solutions based on information about their quality. First, the onlooker bee decides what solution to improve. This is carried out based on the probability of selection as per Equation 2.39. Accordingly, a higher fitness corresponds with a higher probability of being selected.

$$p_i = \frac{fit_i}{\sum_{i=1}^{N_{FS}} fit_i} \qquad (2.39)$$

Once initial data have been accumulated in terms of the probabilities $p_i$ (whose sum equals one), each onlooker bee selects its basic food source by drawing a random number from a uniform distribution between 0 and 1. The onlooker bee then modifies the solution as per Equation (2.38) and decides whether or not to replace the food source. This is carried out in the same manner as applied by the employed bee (described above). Following operation 3, the algorithm memorizes the best best-obtained solution.

The final optimization step (**operation 4**) involves the scout bees, each of which decides whether a potential solution should be further explored or abounded. If the number of attempts $n_{tr}$ to improve a food source quality to make it meet the criteria is equal or higher than $L$, this food source should be replaced by a random food source provided by a scout bee.

Subsequently, the quality of the current food sources is evaluated, and the best-found solution is memorized. The optimization process continues until the criterion of the calculation stop is met.

## 3. Case studies

### 3.1. General inputs

Two different case studies and a sensitivity analysis are carried out to demonstrate the utility of the above-presented approach. The case studies differ in terms of ice conditions and the location of an offshore installation. Both case studies are based on real-world offshore projects with known fleet configurations, which are used as reference fleets (Kjøl 2014), (Staalesen 2019). General input data and assumptions applied in the case studies are presented in Tables 8 and 9.

As per the outlined approach, the optimization algorithm searches for a fleet configuration that minimizes the overall costs, including the expected accident-related costs. A fleet configuration is composed based on a predefined list of candidate vessel types. Specifically, a total of 27 different vessel types are considered, determined based on real-world vessels. In the case studies, the maximum number of vessels of a specific type is assumed to be five, considering their real-world availability.

All considered candidate vessel types and their technical characteristics are listed in the Annex (Table A1 and Table A2). As per industry standard, the following vessel-specific information is provided: year launched, deadweight ($DW$), nominal useful cargo deck area ($S$), power plant capacity ($N_{pp}$), ice class, icebreaking capability ($h_{ice}$), and duty capabilities (i.e. tasks that can be performed by the vessel). Most of the data originate from (Russian Maritime Register of Shipping 2020). All candidate vessel types are given a Fi-Fi and DP classification based on their corresponding capabilities as per Table 3. Other vessel capabilities, including oil recovery, towing, AH, and safety standby, are indicated by Boolean expressions. Specifically, as per Table A1 (see annex), the availability of a function is defined as ¨+¨ (available) or ¨-¨ (not available). It is assumed that all representatives of a specific vessel type have the same characteristics, e.g. year launched.

**Table 8.** Input data for the calculation of vessel charter rates as per Equation 2.33. (Døsen and Langeland 2015).

| Parameter name | Symbol | Value |
|---|---|---|
| Number of days from the contract initiation to the start date of the charter | $DF$ | 180 days |
| Production coefficient (see Equation 2.33) | $K_p$ | 0 |
| Drilling coefficient (see Equation 2.33) | $K_d$ | 0 |
| Brazil coefficient (see Equation 2.33) | $K_{br}$ | 1 |
| Price per barrel of Brent crude oil | $P_{oil}$ | 60 USD |
| Monthly average spot charter rate | $P_{spot}$ | 20000 USD |
| Worldwide monthly oil production | $O_{prod}$ | 400000 m$^3$ |

The optimization process is performed using Microsoft Visual Studio Express, which is an integrated development environment (IDE) application supporting the development of new software. The applied code is written in C# language.

As per the outlined approach, the optimization process is divided into two stages. In the first stage, the 'traditional support fleet' performing supply, AH, towing, safety standby, Fi-Fi, and oil recovery, is optimized. In the second stage, considering the outcome of the first stage, optimization of the 'ice management fleet' is carried out.

In both case studies, the ice conditions at the drilling location are determined based on ice data for July, which represents the period with the most difficult ice conditions during a typical drilling season. Three different winter scenarios are considered: mild, average, and severe. In the optimization, the occurrence of the various scenarios follows the probabilities $F_{ic}$ presented in Table 9. It is noted that sea ice is treated as a risk factor addressed by ice management.

**Table 9.** General input data for the case studies (Døsen and Langeland 2015), (Aas et al. 2007), (Kryzhevich 2017), (RigZone 2020), (Liu et al. 2016), (Russian Maritime Register of Shipping 2014), (Gibson 1999), (Viscusi and Gayer 2002), (Alp maritime services 2020), (Dumanskaya 2014), (Sabodash et al. 2019).

| Parameter name | Symbol | Value |
|---|---|---|
| Monthly cargo consumption rate | $Cons_{rate}$ | 3000 m$^2$ per month |
| Cargo deck area of the installation | $S_{deck}$ | 300 m$^2$ |
| The upper limit value of equivalent ice thickness | $h_{max}$ | 0.7 m |
| Daily charter rate of the installation | $R^{op}$ | 600000 USD |
| Towing speed | $v_{tow}$ | 4 knots |
| Beaufort number | $BN$ | 4 |
| Minimum machinery power for towing operations | $N_{pp, towing}^{min}$ (1,2,3,4 vessels) | 9600, 5530, 4150, 3120 kW |
| Minimum machinery power for AH operations | $N_{pp, AH}^{min}$ | 12000 kW |
| Price of fuel | $p_{fuel}$ | 550 USD |
| Value of a human life | $VH$ | USD 30 million |
| Time spent to support the drilling operations | $t_{op}$ | 90 days |
| Anchor handling time | $t_{AH}$ | 18 days |
| Fuel consumption rate for AH operation | $c_{AH}$ | 30 t/day |
| Fuel consumption rate for cargo operations at the supply port | $c_{v,p}$ | 2 t/day |
| Fuel consumption rate for cargo operations at the installation | $c_{v,pl}$ | 10 t/day |
| Fuel consumption rate for safety standby operation | $c_{SS}$ | 8 t/day |
| Fuel consumption rate during ice management | $c_{IM}$ | 20 t/day |
| Default probability of different ice conditions | $F_{ic}$ (mild, average, severe) | 0.15, 0.7, 0.15 |
| Towing and supply distances (Case study 1) | $dist_{tow}, dist_{sup}$, | 810 |
| Towing and supply distances (Case study 2) | $dist_{tow}, dist_{sup}$, | 770 |
| Probability of iceberg occurrence (Case study 1) | $p_{iceberg}$ | 0.3 |
| Probability of iceberg occurrence (Case study 2) | $p_{iceberg}$ | 0.01 |

Supply vessels operating between the base port and the drilling location are assumed to be able to avoid ice, meaning that they operate continuously in open water. It is also noted that the default ice conditions are determined based on historical data from the period 1939–2012 (Dumanskaya 2014), which does not represent the trend of diminishing ice conditions in the Arctic observed in recent years.

Each case study results in an optimized support fleet configuration. The characteristics and versatility of each optimized fleet composition are analyzed using spider diagrams presenting normalized values of selected key performance indicators (KPIs). Specifically, KPIs are calculated for eight different types of fleet performance: supply, ice management, towing, Fi-Fi, DP class, ice class, fleet age, and environmental friendliness. All KPIs are given a normalized value in the range 1–100. Supply, DP class, ice class, fleet age, and environmental friendliness are normalized as per Table 10. Other KPIs, including ice management, towing, and Fi-Fi, are normalized in accordance with the corresponding applied operational strategy so that Scenario A corresponds to a value of 3/3, Scenario B corresponds to a value of 2/3, and Scenario C corresponds to a value of 1/3. Any normalized KPI value above 100 indicates excess redundancy that cannot be utilized Table 11.

## 3.2. Case study 1

Case study 1 deals with the design of a support fleet for conditions corresponding to the 2014 drilling season at the Universitetskaya-1 well located in the northern Kara Sea. The assumed ice conditions at the location in July are determined as per Table 12.

**Table 10.** Normalization of fleet KPIs.

| KPI | Explanation |
| --- | --- |
| Supply | Equals $100 S_{rate}/Cons_{rate}$. A performance of 100 means that the supply-demand is fully satisfied. |
| DP class | Corresponds to the average DP class of the fleet divided by three (e.g. an average DP class value of 2 corresponds to a DP class performance of 2/3). |
| Ice class | Corresponds to the average ice class of the fleet normalized as per Table 11 (e.g. a vessel consisting of two PC 4 ships and two PC5 ships have an ice class KPI of 68.2). |
| Fleet age | Reflects the average vessel age (assumed to reflect the operational reliability of the vessels, among others). A fleet consisting of brand-new ships would have a KPI value of 100. Each added year of age reduces the KPI value by two (e.g. an average age of 10 years corresponds to a KPI value of 80). |
| Environmental friendliness | Reflects the total amount of fuel used, which is directly proportional to the amount of $CO_2$ produced. A KPI value of 100 corresponds to a reference fuel consumption of 4200 tons. For each additional 165 tons of fuel consumed, the KPI value drops by one. |

**Table 11.** Normalized ice class values.

| Ice class | Normalized value | Ice class | Normalized value |
| --- | --- | --- | --- |
| PC1 | 100 | PC7 | 100·(5/11) |
| PC2 | 100·(10/11) | IA Super | 100·(4/11) |
| PC3 | 100·(9/11) | IA | 100·(3/11) |
| PC4 | 100·(8/11) | IB | 100·(2/11) |
| PC5 | 100·(7/11) | IC | 100·(1/11) |
| PC 6 | 100·(6/11) | Below IC | 0 |

The results of the optimization of the 'traditional support fleet' for Case study 1 are presented in Table 13. The optimized solution for the default ice conditions consists of one vessel of Type 1 with ice class PC 3 (for safety standby), two vessels with ice class PC 4 (for towing, anchor handling, and supply), and two vessels with ice class PC 5 (for supply and firefighting). On the other hand, the reference fleet configuration consists of one vessel with ice class PC 3 (for safety standby) and five vessels with an ice class lower than IC (no ice class) for other duties. The higher number of vessels in the reference fleet can be explained by a preference for high supply capacity redundancy (which is common in the industry) and a higher than average cargo consumption rate. The considerable difference in vessel ice class is likely explained by our conservative assumption concerning the ice conditions as well as by or conservative RIO criterion that excludes operations with elevated operational risk.

Due to climate change, since 2005, the Kara Sea has experienced multiple mild winters (Dumanskaya 2014). In order to consider the effect of the trend of diminishing ice conditions, we repeated the optimization procedure, assuming that all winters are mild and vessels with elevated operational risk (−10 < RIO < 0) are feasible. Based on this assumption, the obtained optimal fleet configuration for the 'traditional support fleet' consists of one vessel of Type 1 with ice class PC 3 (for safety standby), one vessel of Type 10 with ice class PC 4 (for towing, anchor handling, and supply), two vessels of Type 11 with ice class PC 5 (for supply and firefighting), and one vessel of Type 19 with ice class IC (for towing, anchor handling, and supply). It is noted that even under the applied assumptions, vessels with an ice class lower than IC are not considered because of the RIO constraint. The reference fleet dominated by vessels with no ice class is eligible for Case study 1 only if a reliable ice forecast for a specific drilling season indicates no ice in the area of operation, which in the northern

**Table 12.** Default ice conditions for Case study 1 (Dumanskaya 2014), (Shalina and Sandven 2018).

| Scenario | Mild | Average | Severe |
| --- | --- | --- | --- |
| Concentration, $c$ | 0.3 | 0.9 | 1 |
| Level ice thickness, $h_i$[m] | 0.8 | 1.5 | 1.6 |
| Amount of ice ridging, $b$ (see Equation 3.24) | 2 | 2 | 3 |
| Snow thickness, $h_{sn}$[m] | 0.08 | 0.12 | 0.13 |

**Table 13.** Reference and optimized fleet configurations of the 'traditional' support fleet (Case study 1).

| Vessel type | Ice class | Optimized solution 1 (default ice conditions) | Optimized solution 2 (mild ice conditions) | Optimized solution 3 (no ice, supply redundancy) | Reference |
|---|---|---|---|---|---|
| Type 1 | PC 3 | 1 | 1 | 0 | 1 |
| Type 10 | PC 4 | 2 | 1 | 0 | 0 |
| Type 11 | PC 5 | 2 | 2 | 0 | 0 |
| Type 16 | No ice class | 0 | 0 | 0 | 2 |
| Type 17 | No ice class | 0 | 0 | 0 | 2 |
| Type 18 | No ice class | 0 | 0 | 2 | 0 |
| Type 19 | IC | 0 | 1 | 0 | 0 |
| Type 23 | No ice class | 0 | 0 | 1 | 0 |
| Type 27 | No ice class | 0 | 0 | 3 | 1 |

Kara Sea is possible during a year with easy ice conditions.

To better demonstrate the developed method's merits, we repeated the optimization process using assumptions similar to those used in practice for the real reference fleet, namely, ice is not considered, and a specific supply redundancy is required. Some level of supply redundancy is often considered in practice to avoid unnecessary operational downtime. The supply redundancy of the reference fleet is estimated at 17%, meaning that the Supply KPI of the optimized fleet must be 117 or more (see Table 10). Based on these assumptions, the obtained optimal fleet has no ice class and consists of one vessel of Type 23 (for safety standby), two vessels of Type 18 (for towing, anchor handling, and supply), and three vessels of Type 27 (for supply and firefighting). The obtained fleet is identical to the reference fleet in terms of ice class and number of vessels.

The ice management fleet optimization results for Case study 1, assuming default ice conditions and mild ice conditions (eligible RIO > −10), are presented in Table 14. The reference fleet and two optimized fleets consist of five vessels, including a lead vessel with ice class PC 3, representing some minor deviation from the complete ice management scenario (see Table 7). However, the ice classes of the secondary vessels are different: the fleet optimized for the default ice conditions has the highest icebreaking capabilities (secondary vessels of ice class PC 5); the fleet optimized for the mild ice conditions has the lowest icebreaking capabilities (secondary vessels of ice class IB); the reference fleet is a tradeoff between the derived optimal solutions (secondary vessels of ice class IA – PC 4). The ice management fleet optimization cannot be provided for the case of no ice (optimized solution 3), as ice management would require ice to be reasonable. However, for further analysis, the ice management fleet for optimized solution 3 is conservatively assumed to be the same as that of optimized solution 2 (mild ice conditions) to compare the fleets consistently.

Further analysis of Case study 1 is presented for the complete support fleet (traditional + ice management). The four analyzed fleet configurations (see Tables 13 and 14) differ in terms of functionality

(KPIs) and total cost. As per the spider diagrams presented in Figures 1 and 2, the first optimized solution (default ice conditions) performs better in terms of 'Fi-Fi', 'Ice class', and 'Environmental friendliness' compared to the reference solution, whereas it performs worse in terms of 'Supply' and 'Fleet age'. As per the spider diagrams presented in Figure 2 and Figure 3, the second optimized fleet (mild ice conditions) is close to the reference fleet in KPIs but has an advantage in terms 'Fi-Fi', 'Ice class', whereas it performs worse in terms of 'Supply' and 'Fleet age'. The spider diagram in Figure 4 is identical to the spider diagram presented in Figure 2, indicating that the optimized solution 3 and the reference fleet are equal in terms of functionality.

The total cost of the reference fleet is estimated at around USD 55.1 million per drilling season, whereas the corresponding costs of the optimized fleet for the default ice conditions and the mild ice conditions are estimated at around USD 49.3 million (10% lower costs than the reference fleet) and USD 46.0 million per drilling season (16% lower costs than the reference fleet), respectively. Furthermore, the total cost estimate for the optimized fleet obtained assuming supply redundancy and ice free conditions is estimated at USD 48.9 million per drilling season, i.e. 11% lower than for the reference fleet although it provides the same functionality. This indicates that the optimized solutions provide enhanced cost-efficiency. The presented KPIs spider diagrams, together with the cost estimations, provide valuable information for decision making while they help to find a tradeoff between fleet performance and costs. Figure 5 presents the contributions of different cost categories to the objective function for optimized solution 1. The individual cost contributions of the vessels are not presented because the influence of individual vessels is different for different fleet configurations. Thus, it is only motivated to study the efficiency of the vessels as a part of a support system.

Figure 6 shows the ABC-based optimization progress for Case study 1, traditional support fleet (default ice conditions scenario). The starting point of the optimization process is chosen randomly by the ABC algorithm because, in real life, there is typically no relevant reference fleet to consider. The fleet composition corresponding to each iteration is presented in Table A3. The algorithm reached the optimal solution in 21 iterations, which is efficient considering that the total number of potential solutions is $6^{11} = 362,797,056$, as calculated based on the number of possible combinations per vessel type (from 0 to 5) and the number of feasible vessels that satisfy the constraints (11).

**Table 14.** Reference and optimized fleet configurations of the ice management support fleet (Case study 1).

| Vessel type | Ice class | Optimized solution (default ice conditions) | Optimized solution 2 (mild ice conditions) | Optimized solution 3 (no ice, supply redundancy), assumed fleet | Reference |
|---|---|---|---|---|---|
| Type 1 | PC 3 | 1 | 1 | 1 | 1 |
| Type 7 | PC 5 | 4 | 0 | 0 | 0 |
| Type 10 | PC 4 | 0 | 0 | 0 | 1 |
| Type 12 | IA Super | 0 | 0 | 0 | 2 |
| Type 13 | IA | 0 | 0 | 0 | 1 |
| Type 15 | IB | 0 | 4 | 4 | 0 |

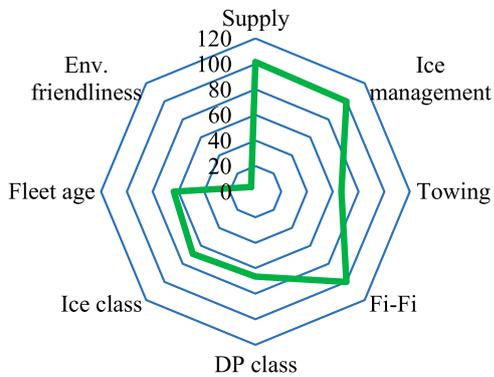

**Figure 1.** Spider diagram for the optimized fleet (Case study 1, default ice conditions scenario).

### 3.3. Case study 2

Case study 2 deals with the design of a support fleet for conditions corresponding to the 2017 drilling season at the Leningradskoe oil field located in the southern Kara Sea. The assumed ice conditions at the location in July are specified as per Table 15, which are milder than those of Case study 1.

As per Table 16, the calculated optimal fleet configuration of the traditional support fleet for the default ice conditions is the same as for Case study 1 (default ice conditions), which is related to the RIO constraint (see Equation 2.5) calculated based on the severe ice conditions scenario. The reference fleet configuration consists of one vessel with ice class PC 3 (for safety standby) and five vessels with an ice class lower than IC (no ice class) for other duties, which is justified by the ice conditions of the 2017 year at the Leningradskoe oil field (mild ice conditions scenario).

We repeated the optimization procedure for the mild ice condition scenario. The obtained optimal fleet configurations for the traditional support fleet consists of a single vessel with no ice class (for safety standby) and four vessels with an ice class below IC: one of Type 16 (for supply and firefighting), two of Type 18 (for towing, anchor handling, and supply), and one of Type 27 (for supply and firefighting). This solution is close to the corresponding reference solution in terms of ice class (see Table 16).

The optimization is repeated for Case 2 to provide the same level of supply redundancy for the optimal solution and the reference fleet, estimated at around 20% (this indicates that the Supply KPI of the optimized fleet must be at least 120 as per Table 10). The obtained fleet (Table 16) has no ice class and consists of a single vessel of Type 23 (for safety standby), three vessels of Type 18 with (for towing, anchor handling, and supply), and two vessels of Type 27 (for supply and firefighting). The obtained fleet is close to the reference fleet in terms of ice class and number of vessels.

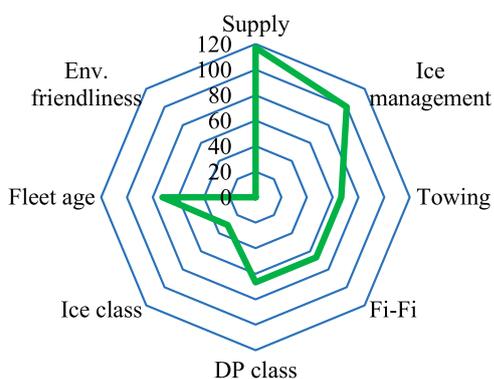

**Figure 2.** Spider diagram for the reference fleet configuration (Case study 1).

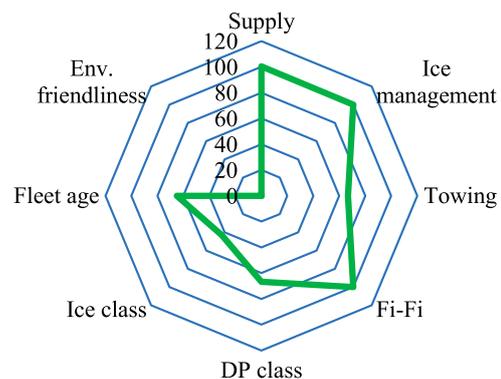

**Figure 3.** Spider diagram for the optimized fleet (Case study 1, mild ice conditions scenario).

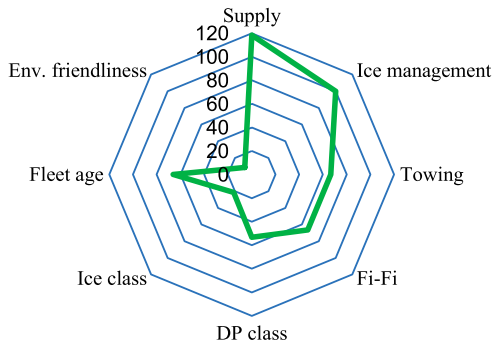

**Figure 4.** Spider diagram for the optimized fleet (Case study 1, no ice, supply redundancy).

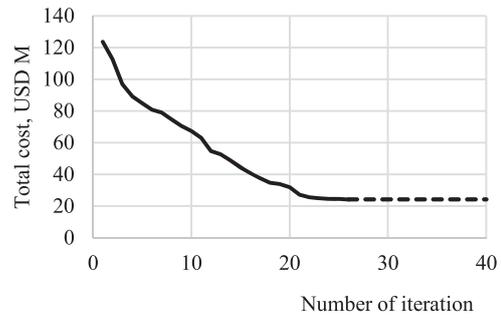

**Figure 6.** Progress of the ABC-based optimization for traditional support fleet (Case study 1, optimized solution 1).

Regarding the 'ice management' support fleet, as per Table 17, the reference and three optimized fleet configurations consist of 1 vessel with ice class PC 3. This solution represents the passive ice management strategy (see Table 7).

Further analysis of Case study 2 is presented for the complete support fleet (traditional + ice management). As per the spider diagrams presented in Figures 7 and 8, the optimized solution (default ice conditions) performs better in terms of 'Fi-Fi' and 'ice class', whereas the reference solution performs better in terms of 'supply' and 'fleet age'. The solution optimized for mild winters (Figure 9) has a similar spider diagram shape as the reference solution indicating similar functionality, and shows the best performance in terms of environmental friendliness. As per Figure 8 and Figure 10, in terms of supply redundancy the performance of the reference fleet and the optimized fleet is identical.

The total cost for the reference fleet is estimated at around USD 30.2 million. The estimated total cost of the fleet optimized for the default ice conditions is similar at USD 29.4 million. Figure 11 demonstrates the contribution of different costs to the objective function for optimized solution 1. The estimated total cost of the fleet optimized for exclusively 'mild' winters is around 25% lower at USD 22.1 million. The total cost for the fleet with the same functionality as the reference fleet (optimized solution 3) optimized for the mild ice condition is estimated at USD 24.5 million, or around 19% lower than the costs of the reference fleet.

Case study 2 shows that the application of the proposed optimization approach could result in substantial money savings, assuming that reliable input data, e.g. ice condition forecasts, is provided. Figure 12 demonstrates the ABC-based optimization progress for Case study 2, traditional support fleet (mild ice conditions scenario). The algorithm reached the optimal solution in 29 iterations, which is efficient considering that the total number of potential solutions is $6^{16} = 282,110,9907456$, as calculated based on the number of possible combinations per one vessel type (from 0 to 5) and the number of feasible vessels that satisfy the constraints (16).

### 3.4. Sensitivity analysis

A sensitivity analysis is performed separately for the traditional support fleet and the ice management fleet. The sensitivity analysis for the traditional support fleet is carried out by varying the value of individual input parameters within a specific range determined based on an assumed minimum and maximum feasible value of each parameter (see Table 18). The feasible range of each parameter variation is set as a specific percentage of the default value. The applied percentage range is different for different parameters to reflect their real-life values. We assume that all default values for the input parameters are the same as applied in Case study 1–2 (see Tables 8 and 9), except for the distances ($dist_{tow}$, $dist_{sup}$) that are both assumed to be 200 NM, and the price of fuel ($p_{fuel}$) that is assumed to be 500 USD/ton. As the impact of sea ice is already studied in Case study 1–2, we do not consider sea ice in the sensitivity analysis for the traditional support fleet. This means any of the listed candidate vessel types (Table A1) can be considered for the traditional support fleet.

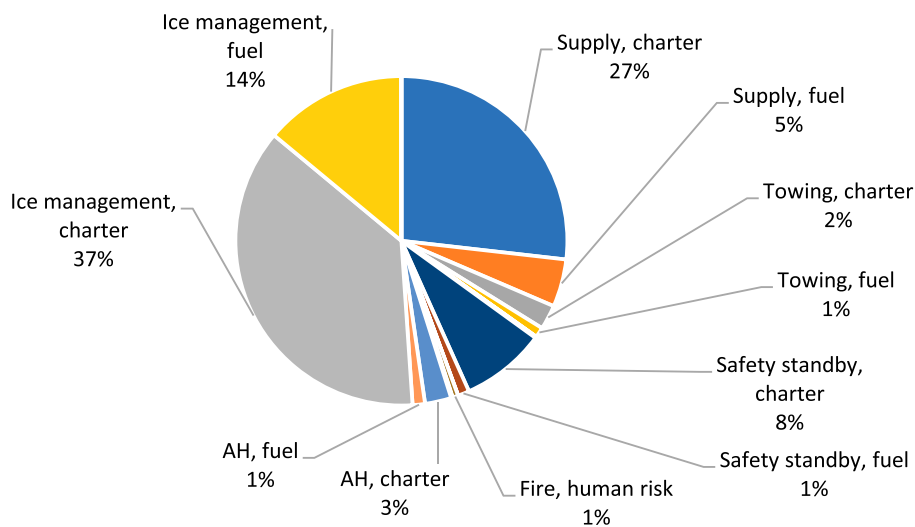

**Figure 5.** Contribution of different cost categories to the objective function, Case study 1, optimized solution 1 (default ice conditions).

**Table 15.** Ice condition parameters for Case study 2 (Dumanskaya [2014]), (Shalina and Sandven [2018]).

| Scenario | Mild | Average | Severe |
|---|---|---|---|
| $c$ | 0 | 0.3 | 1 |
| $h_i$ | 0 | 0.9 | 1.3 |
| $B$ | 0 | 2 | 2 |
| $h_{sn}$ | 0 | 0.08 | 0.11 |

The outcome of the sensitivity analysis is presented in Table 18 and Table 19. Thirty optimization experiments resulted in nine original fleet configurations (solutions S1 – S9) composed of vessels of eight different types from the complete list of 27 vessel types. The same optimal solutions often occurred in different parts of Table 18, meaning that some fleet configurations and vessel types have robust efficiency under varied circumstances, which may be due to the applied constraints.

Table 20 presents relative changes in the expected total cost for all optimization experiments as a percentage of that of the default optimal solution. We find that the optimization outcomes are sensitive to variations in the monthly cargo consumption rate ($Cons_{rate}$), the price of fuel ($p_{fuel}$) and the installation's cargo deck area ($S_{deck}$), whereas they are insensitive to variations in the daily charter rate ($CR_v$) and time spent to support the drilling operations ($t_{op}$). It is noted that $CR_v$ is varied equally (in percent) for all the vessels considered. As per Table 20, the expected total cost of the optimal fleet is the most sensitive to variations in $CR_v$ and $t_{op}$. It noted that some parameter variations may result in a change in the optimal fleet configuration solution but only result in a minor change in costs, whereas other parameter variations may not result in a change in the fleet configuration but result in a significant change in costs.

It is noted that both the cargo deck area of the installation ($S_{deck}$) and distance ($dist_{tow}$, $dist_{sup}$) appear to have a nonlinear impact on the estimated total costs. This means that simultaneous variations in these parameters may significantly influence the performance of a fleet, as well as the overall costs. It is noted that the impact of $S_{deck}$ variation is highly sensitive to the assumed available types of candidate vessels: low $S_{deck}$ may significantly deteriorate the performance of large vessels.

As per Table 18, we find that variations in the risk-related parameters, the value of human life ($VH$), and asset loss ($E_d^{ra}$), have a significant impact on the optimal configuration of the fleet. For instance, as per Table 18, variations in the assumed values of $VH$ and $E_d^{ra}$ between 0–200% results in different optimal fleet configurations (S2, S1, S3). As per Figure 13, configuration S3 has a significantly higher level of safety than S2. This indicates that investing in additional safety might be cost-effective.

The results of the sensitivity analysis of the ice management fleet are presented in Table 21. The different ice conditions (mild, average, severe) are determined as the average of the corresponding ice conditions defined for case study 1–2 in Tables 12 and 15. The probability of iceberg occurrence ($p_{iceberg}$) is assumed to be zero for better interpretability of results.

As per Table 21, the optimal configuration of the ice management fleet (I1 – I3) depends on the assumed scenario. The various ice management fleet configuration solutions, defined as per Table 22, represent each of the considered ice management strategies. In scenario 1 (default scenario), the solution corresponds to the 'complete' ice management strategy. In scenario 2, in which the assumed frequency of occurrence of severe ice conditions is reduced by 5% in favour of average ice conditions, the optimal solution corresponds to the 'active' ice management scenario. In scenario 3, where the assumed ice conditions are further decreased, the obtained fleet configuration corresponds to the 'passive' ice management strategy. In scenarios 4–5, the reduced operational cost of the drilling installation $E_{inst}^{op}$ resulted in the optimal solution I2, which corresponds to the 'active' ice management scenario.

**Table 16.** Reference and optimized fleet configurations of the 'traditional' support fleet (Case study 2).

| Vessel type | Ice class | Optimized solution 1 (default ice conditions) | Optimized solution 2 (mild ice conditions) | Optimized solution 3 (mild ice conditions, supply redundancy) | Reference |
|---|---|---|---|---|---|
| Type 1 | PC 3 | 1 | 0 | 0 | 1 |
| Type 10 | PC4 | 2 | 0 | 0 | 0 |
| Type 11 | PC5 | 2 | 0 | 0 | 0 |
| Type 16 | No ice class | 0 | 1 | 0 | 2 |
| Type 18 | No ice class | 0 | 2 | 3 | 0 |
| Type 23 | No ice class | 0 | 1 | 1 | |
| Type 26 | No ice class | 0 | 0 | 0 | 3 |
| Type 27 | No ice class | 0 | 1 | 2 | 0 |

**Table 17.** Reference and optimized fleet configurations of the 'ice management' support fleet (Case study 2).

| Vessel type | Ice class | Reference | Optimized solution 1 (default ice conditions) | Optimized solution 2 (mild ice conditions) | Optimized solution 3 (mild ice conditions, supply redundancy) |
|---|---|---|---|---|---|
| Type 1 | PC3 | 1 | 1 | 1 | 1 |

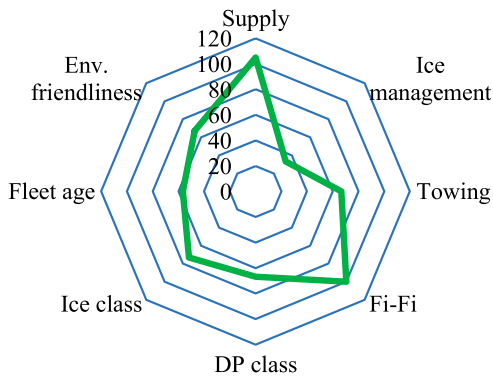

**Figure 7.** Spider diagram for the optimized fleet (Case study 2, default ice conditions scenario).

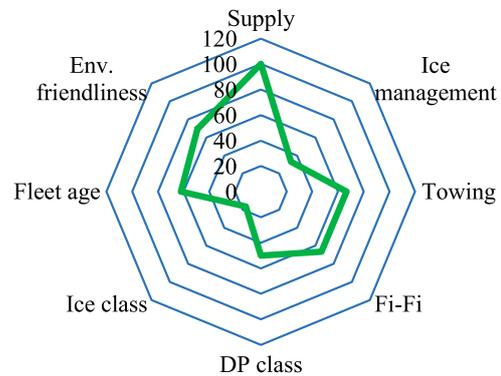

**Figure 9.** Spider diagram for the optimized fleet (Case study 2, mild ice conditions scenario).

## 4. Discussion and conclusions

This article presents an optimization-based approach for sizing and composition of an Arctic offshore drilling support fleet considering cost-efficiency. The approach supports a quantitative assessment of a fleet's functionality, considering the combined effect of: (a) the expected costs of accidental events, (b) the versatility of individual support vessels, and (c) ice management. The considered expected costs include equivalent costs related to human injuries and loss of life, as well as direct economic losses due to production downtime and asset damages.

The proposed optimization process demonstrated a systematic and objective search for a support fleet configuration that minimizes the overall expected costs while meeting set system criteria. Two case studies and a sensitivity analysis were carried out to demonstrate the utility of the approach. The obtained solutions were found to be similar to corresponding real-life support fleet solutions, indicating that the approach works in principle. Conducted case studies demonstrated a significant potential of the approach in reducing fleet operation cost. Specifically, in relation to the reference solutions, the obtained savings were up to 25%.

A performance overview is provided for each optimized fleet composition in terms of a spider diagram presenting different key performance indicators (KPIs). If the approach is used to produce a set of candidate solutions optimized for different assumptions concerning the operating conditions, the spider diagrams may facilitate the final choice of a solution.

The studied optimization problem is highly constrained to ensure the minimum acceptable fleet functionality and safety in specific circumstances. Case studies 1–2 demonstrate that the size of the feasible design space, or the number of possible solutions, significantly depends on the assumed ice conditions. Although the applied constraints affect the optimization process much, they are not the only ones defining the optimal solution: the optimization results are obtained through the actual minimization of the objective (the estimated total cost). That conclusion is based on the sensitivity analysis and data provided in Table A3 for one of the most constrained case study considered (case study 1, default ice conditions). As per Table A3, the optimal fleet is obtained using step-by-step objective improvement considering different combinations of vessels.

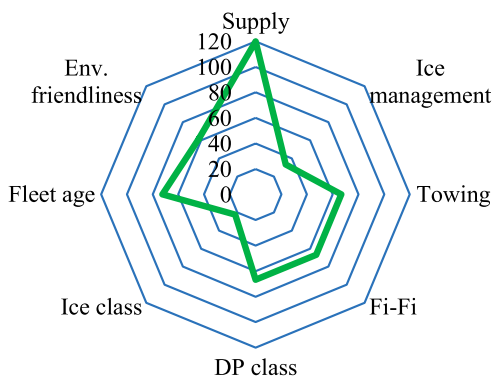

**Figure 8.** Spider diagram for the reference fleet (Case study 2).

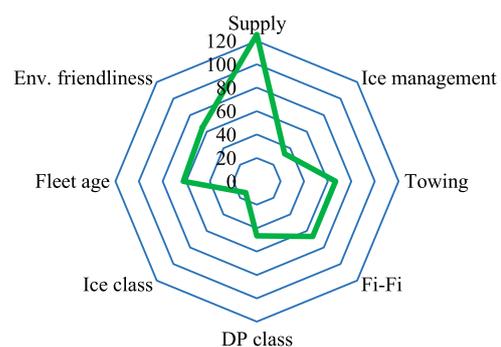

**Figure 10.** Spider diagram for the optimized fleet (Case study 2, mild ice conditions scenario, supply redundancy).

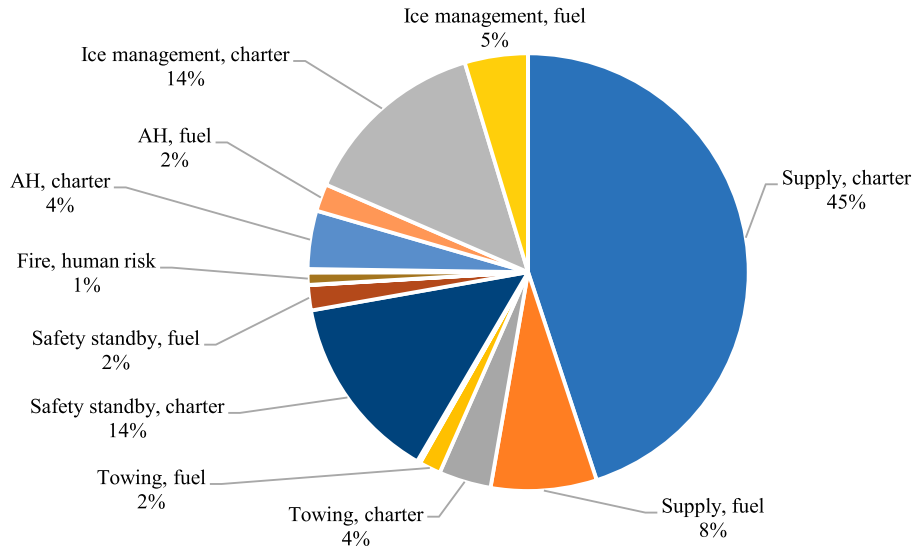

**Figure 11.** Contribution of different cost categories to the objective function (case study 2, optimized solution 1, default ice conditions).

Conducted sensitivity analyses indicate the following. First, the consideration of risk-related costs has a significant impact on the obtained optimal solution. Specifically, we find that it is economically motivated to invest in a fleet with good safety performance in terms of towing, firefighting, dynamic positioning, and ice class. Second, the optimal fleet configuration and the related cost level are sensitive to the assumed ice conditions. Therefore, the prediction and modelling of ice conditions are essential.

Third, some parameter variations (e.g. the daily charter rate, time spent to support the drilling operations) may have a significant effect on the total costs but a limited effect on the optimal fleet configuration. As per Figures 5 and 11, for the default input, charter costs is the most significant cost category. Therefore, a further increase in charter costs affects the optimization results insignificantly. A significant decrease in charter rates, on the other hand, would make other factors more important, resulting in a different optimal solution. The time spent to support the drilling operations equally affects the most important components of the total cost – the fleet charter and fuel costs. As a result, time increasing up to 150% does not change the optimal fleet. Further increasing the time on supporting the operations improves the importance of other factors (e.g. the role of risk-related components), resulting in a different optimal solution.

Fourth, some parameters (e.g. the cargo deck area of the installation and voyage distance) appear to have a nonlinear effect on the estimated total costs. For the cargo deck area of the installation, this effect is related to Equation 2.17 as follows: if the cargo deck area of the installation is higher than the maximum cargo deck area of the candidate vessels, the impact on the estimated total costs is constant. For the voyage distance, the nonlinear effect may be associated with the complex interaction of the applied constraints and algorithms.

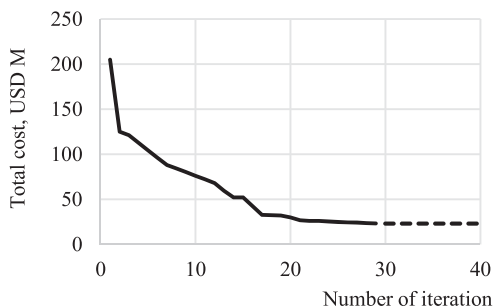

**Figure 12.** Progress of the ABC-based optimization (Case study 2, traditional support fleet, mild ice conditions scenario).

**Table 18.** Optimal fleet configuration solutions (S) for various input parameters. The table determines the impact of variations in individual parameter values on the optimization results (marked from green to grey – from the most sensitive to the least sensitive parameter).

| Value, % | 0 | 50 | 100 | 150 | 200 | 400 |
|---|---|---|---|---|---|---|
| $VH$ and $E_d^{ra}$ | S2 | S1 | S1 | S3 | S3 | – |
| $p_{fuel}$ | – | S3 | S1 | S1 | S4 | – |
| $CR_v$ | – | S4 | S1 | S1 | S1 | – |
| $Cons_{rate}$ | – | S5 | S1 | S6 | S7 | – |
| $t_{op}$ | – | S1 | S1 | S1 | S4 | – |
| $dist_{tow}, dist_{sup}$ | – | S5 | S1 | S1 | S1 | S8 |
| $S_{deck}$ | – | S9 | S1 | S5 | S5 | – |

**Table 19.** Optimal fleet configuration solutions (S) for the 'traditional' support fleet.

| Vessel type | S1 | S2 | S3 | S4 | S5 | S6 | S7 | S8 | S9 |
|---|---|---|---|---|---|---|---|---|---|
| Type 1 | 1 | 1 | 1 | 1 | 1 | 1 | 1 | 1 | 1 |
| Type 11 | 0 | 0 | 0 | 2 | 0 | 0 | 2 | 1 | 0 |
| Type 15 | 0 | 0 | 0 | 0 | 0 | 0 | 0 | 0 | 1 |
| Type 16 | 1 | 0 | 2 | 0 | 2 | 1 | 0 | 1 | 2 |
| Type 18 | 1 | 1 | 0 | 0 | 0 | 1 | 2 | 2 | 0 |
| Type 20 | 0 | 1 | 1 | 0 | 0 | 1 | 0 | 0 | 1 |
| Type 23 | 0 | 1 | 0 | 0 | 0 | 0 | 0 | 0 | 0 |
| Type 27 | 1 | 0 | 0 | 0 | 0 | 1 | 0 | 0 | 0 |

**Table 20.** Percentual (%) change in the expected total cost due to a variation in a single parameter.

| Value, % | 0 | 50 | 100 | 150 | 200 | 400 |
|---|---|---|---|---|---|---|
| $VH$ and $E_d^{ro}$ | −14 | −2 | 0 | +1 | +3 | – |
| $p_{fuel}$ | – | −9 | 0 | +9 | +18 | – |
| $CR_v$ | – | −40 | 0 | +39 | +78 | – |
| $Cons_{rate}$ | – | −12 | 0 | +13 | +25 | – |
| $t_{op}$ | – | −41 | 0 | +41 | +81 | – |
| $dist_{tow}$ $dist_{sup}$ | – | −15 | 0 | +2 | +5 | +32 |
| $S_{deck}$ | – | +16 | 0 | −12 | −12 | – |

**Table 21.** Optimization results for ice management sensitivity analysis.

| Nr | Scenario | Solution | Total cost change |
|---|---|---|---|
| 1 | Default $F_{ic}$, $E_{inst}^{op}$ 100% | I1 | 0.00 |
| 2 | $F_{ic}$ (mild 0.15, average 0.75, severe 0.1) | I2 | −10% |
| 3 | $F_{ic}$ (mild 0.875, average 0.025, severe 0.1) | I3 | −15% |
| 4 | $E_{inst}^{op}$ 50% | I2 | −11% |
| 5 | $E_{inst}^{op}$ 0% | I2 | −23% |

**Table 22.** Calculated optimal configurations of the 'ice management' fleet.

| Vessel type | Ice class | I1 | I2 | I3 |
|---|---|---|---|---|
| Type 1 | PC 3 | 1 | 1 | 1 |
| Type 7 | PC 5 | 4 | 1 | 0 |

As shown in the case studies, the metaheuristic ABC algorithm demonstrated high performance and efficiency in the optimization of an Arctic offshore drilling support fleet with a significant number of candidate vessel types. Despite the metaheuristic and stochastic nature of the applied ABC optimization algorithm, separate optimization rounds resulted in the same solution, indicating that the approach is robust.

Future studies are recommended to address the limitations of the present approach include the following. First, the provided scenario-based models for the quantitative assessment of a fleet functionality are discrete, meaning that an integer value determines each state. These models could be independently evolved into continuous algorithms that would significantly improve the validity and utility of the method. Second, stochastic factors (e.g. stochastic weather factors) are not included in the model. Some stochastic entities (e.g. the weather influence on voyage times) are relevant to assess the robustness of supply routing and scheduling but assumed to have a limited impact on the optimal support fleet configuration (Ehlers et al. 2019). However, weather-related factors might influence, e.g. the operating windows for specific cargo operations. Also, the combined impact of different stochastic factors might be significant. Third, the approach could be expanded and applied to year-round offshore production. This would require the incorporation of methodologies for the calculation of a vessel's fuel consumption and speed in the ice of various thicknesses, as well as the consideration of hull ice loading. The consequences of potential ice damages to ships and installations are not considered. The impact of this limitation is assumed insignificant for the case of exploration drilling, both because the considered operations occur in summer when the occurrence of sea ice is rare, and because ice management is provided. Fourth, an additional investigation of the role of logistics factors (e.g. using a floating storage) on the efficiency and safety of an Arctic offshore support fleet could be carried out. Finally, additional cases, such as the use of supply hovercrafts in shallow water conditions with thick ice, could be investigated.

## Acknowledgments

We would like to thank Prof. Dr. Jani Romanoff and Dr. Alex Topaj for their valuable comments and suggestions to improve the article. This project has received funding from the Lloyd's Register Foundation, a charitable foundation, helping protect life and property by supporting engineering-related education, public engagement, and the application of research www.lrfoundation.org.uk.

## Disclosure statement

No potential conflict of interest was reported by the author(s).

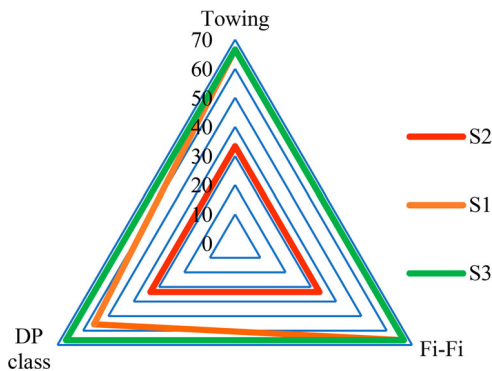

**Figure 13.** Safety-related KPIs for fleet configuration S1 – S3.

## Funding

This work was supported by Lloyd's Register Foundation.

## ORCID

*A. A. Kondratenko*  http://orcid.org/0000-0003-3008-4498
*M. Bergström*  http://orcid.org/0000-0001-7758-3038
*M. Suominen*  http://orcid.org/0000-0001-9758-9365
*P. Kujala*  http://orcid.org/0000-0003-2665-9957

Table A1. General characteristics of different support vessel types.

| Name | Year of delivery | $DW$ [ton] | $S$ [m$^2$] | $N_{pp}$[MW] | Ice class | $h_{ice}$ [m] | Fi-Fi class | DP class | Oil recovery | Towing | Anchor handling | Standby |
|---|---|---|---|---|---|---|---|---|---|---|---|---|
| Type 1 (Vladimir Ignatiuk) | 1982 | 2110 | 470 | 17 | PC 3 | 1.8 | 0 | 0 | − | + | + | + |
| Type 2 (Toboy) | 2008 | 1930 | 490 | 13.55 | PC 4 | 1.5 | 0 | 0 | − | − | − | − |
| Type 3 (SCF Sakhalin) | 2005 | 4300 | 700 | 17.4 | PC 4 | 1.5 | 3 | 1 | + | − | − | + |
| Type 4 (Yury Topchev) | 2005 | 3860 | 750 | 20 | PC 4 | 1.7 | 1 | 1 | + | + | − | + |
| Type 5 (Vitus Bering) | 2012 | 4160 | 710 | 18 | PC 4 | 1.5 | 3 | 2 | + | + | − | + |
| Type 6 (SCF Endurance) | 2006 | 4370 | 780 | 14 | PC 5 | 1.3 | 0 | 2 | + | − | − | + |
| Type 7 (Polar Pevek) | 2006 | 1300 | 300 | 8.2 | PC 5 | 1.2 | 3 | 0 | + | + | − | − |
| Type 8 (Gennadiy Nevelskoy) | 2017 | 3260 | 500 | 20 | PC 4 | 1.5 | 3 | 2 | + | − | − | + |
| Type 9 (Stepan Makarov) | 2017 | 3670 | 500 | 20 | PC 4 | 1.5 | 3 | 2 | + | − | − | + |
| Type 10 (CCGS Vincent Massey) | 2000 | 3000 | 600 | 13.4 | PC4 | 1.2 | 0 | 2 | − | + | + | + |
| Type 11 (Aleut) | 2015 | 2600 | 600 | 14 | PC 5 | 1.2 | 3 | 2 | + | + | + | − |
| Type 12 (Brage Viking) | 2012 | 4500 | 750 | 14 | IA Super | 1 | 2 | 2 | + | + | + | + |
| Type 13 (Magne Viking) | 2011 | 4500 | 750 | 14 | IA | 0.9 | 2 | 2 | + | + | + | + |
| Type 14 (Kigoriak) | 1979 | 1690 | 520 | 12.3 | IA Super | 0.91 | 0 | 0 | − | + | + | − |
| Type 15 (Svetlyy) | 2009 | 890 | 250 | 5.3 | IB | 0.7 | 0 | 1 | + | + | − | + |
| Type 16 (Siem Amethyst) | 2009 | 4250 | 810 | 20 | 0 | 0.15 | 2 | 2 | + | + | + | + |
| Type 17 (Normand Supporter) | 2010 | 5300 | 1075 | 7.5 | 0 | 0.15 | 0 | 2 | + | − | − | − |
| Type 18 (Maersk Handler) | 2002 | 2620 | 520 | 12.9 | 0 | 0.15 | 0 | 1 | + | + | + | − |
| Type 19 (Maersk Dispatcher) | 2005 | 4050 | 755 | 13.4 | IC | 0.4 | 1 | 2 | − | + | + | − |
| Type 20 (Bourbon Liberty 151) | 2013 | 1700 | 400 | 3.7 | 0 | 0 | 1 | 2 | + | − | − | − |
| Type 21 (Havila Foresight) | 2008 | 4780 | 1050 | 8.8 | 0 | 0 | 0 | 2 | + | − | − | − |
| Type 22 (Normand Skipper) | 2005 | 6400 | 1220 | 8.4 | 0 | 0 | 0 | 2 | + | − | − | + |
| Type 23 (VOS Triton) | 2006 | 1740 | 340 | 3.8 | 0 | 0 | 1 | 0 | − | + | + | + |
| Type 24 (Energy Scout) | 2007 | 3300 | 690 | 4 | 0 | 0 | 1 | 2 | + | − | − | − |
| Type 25 (Normand Prosper) | 2010 | 5000 | 760 | 24.7 | 0 | 0 | 0 | 2 | − | + | + | − |
| Type 26 (Sea Spear) | 2014 | 4450 | 1000 | 6.5 | 0 | 0 | 1 | 2 | + | − | − | − |
| Type 27 (Siem Pilot) | 2010 | 5000 | 930 | 8.4 | 0 | 0 | 2 | 2 | + | + | − | + |

**Table A2.** Estimated daily time charter rates, cruising fuel consumption rates, and cruising speeds.

| Name | Displacement [t] | $C_b$ | $L_{pp}$[m] | $B$ [m] | $D$ [m] | $d$ [m] | Charter rate, USD/day | $c_{v,mov}$[ton/day] | Cruising speed [knot] |
|---|---|---|---|---|---|---|---|---|---|
| Type 1 (Vladimir Ignatiuk) | 7080 | 0.60 | 79.2 | 17.5 | 10 | 8.3 | 32,750 | 15.1 | 8.80 |
| Type 2 (Toboy) | 6530 | 0.55 | 73.3 | 18 | 11.2 | 8.8 | 35,450 | 16.8 | 8.98 |
| Type 3 (SCF Sakhalin) | 9980 | 0.66 | 93.4 | 21.2 | 11 | 7.5 | 40,020 | 18.8 | 8.79 |
| Type 4 (Yury Topchev) | 9670 | 0.74 | 84.4 | 19 | 10.5 | 8 | 43,425 | 17.9 | 8.72 |
| Type 5 (Vitus Bering) | 10,710 | 0.68 | 91.4 | 21.2 | 11 | 7.9 | 44,670 | 19.6 | 8.79 |
| Type 6 (SCF Endurance) | 8910 | 0.71 | 77.6 | 19 | 10 | 8.25 | 42,160 | 18.4 | 8.73 |
| Type 7 (Polar Pevek) | 4570 | 0.67 | 64.5 | 17 | 8 | 6.1 | 28,510 | 10.7 | 8.50 |
| Type 8 (Gennadiy Nevelskoy) | 10,710 | 0.68 | 91.4 | 21.2 | 11 | 7.9 | 40,650 | 19.6 | 8.79 |
| Type 9 (Stepan Makarov) | 10,960 | 0.68 | 93.91 | 21.2 | 11 | 7.9 | 40,820 | 19.8 | 8.79 |
| Type 10 (CCGS Vincent Massey) | 6870 | 0.69 | 75.2 | 18 | 8.5 | 7.2 | 35,160 | 15.0 | 8.61 |
| Type 11 (Aleut) | 7970 | 0.72 | 76.8 | 19.5 | 8.5 | 7.25 | 38,340 | 15.8 | 8.66 |
| Type 12 (Brage Viking) | 9450 | 0.724 | 76.2 | 22 | 9 | 7.6 | 38,910 | 15.6 | 8.77 |
| Type 13 (Magne Viking) | 9400 | 0.720 | 76.2 | 22 | 9 | 7.6 | 37,305 | 15.6 | 8.77 |
| Type 14 (Kigoriak) | 6850 | 0.60 | 78.9 | 17 | 10 | 8.3 | 24,820 | 17.0 | 8.79 |
| Type 15 (Svetlyy) | 2700 | 0.65 | 62.4 | 15 | 6.2 | 4.3 | 23,470 | 7.4 | 8.11 |
| Type 16 (Siem Amethyst) | 10,800 | 0.754 | 79.35 | 22 | 9.6 | 8 | 27,520 | 17.6 | 8.82 |
| Type 17 (Normand Supporter) | 8640 | 0.73 | 84.9 | 20 | 8.3 | 6.8 | 32,355 | 14.3 | 8.64 |
| Type 18 (Maersk Handler) | 5910 | 0.68 | 69.3 | 18 | 8 | 6.8 | 20,760 | 11.8 | 8.54 |
| Type 19 (Maersk Dispatcher) | 9420 | 0.73 | 76.2 | 21.9 | 9 | 7.5 | 29,310 | 15.5 | 8.77 |
| Type 20 (Bourbon Liberty 151) | 3010 | 0.73 | 58.5 | 14 | 5.8 | 4.9 | 20,250 | 11.6 | 8.14 |
| Type 21 (Havila Foresight) | 8040 | 0.73 | 86.6 | 19.7 | 7.85 | 6.3 | 28,980 | 14.9 | 8.58 |
| Type 22 (Normand Skipper) | 10,050 | 0.74 | 84.3 | 22 | 8.6 | 7.1 | 31,760 | 17.6 | 8.74 |
| Type 23 (VOS Triton) | 2770 | 0.69 | 52.2 | 15 | 6.1 | 5 | 17,630 | 15.8 | 8.27 |
| Type 24 (Energy Scout) | 4880 | 0.768 | 66.8 | 16 | 7 | 5.8 | 23,280 | 11.4 | 8.42 |
| Type 25 (Normand Prosper) | 11,410 | 0.70 | 84.8 | 24 | 9.8 | 7.8 | 25,070 | 18.3 | 8.81 |
| Type 26 (Sea Spear) | 7700 | 0.73 | 82 | 19 | 8 | 6.6 | 29,130 | 14.2 | 8.59 |
| Type 27 (Siem Pilot) | 8460 | 0.745 | 77.2 | 20 | 8.6 | 7.15 | 27,520 | 15.5 | 8.70 |

**Table A3.** Fleet configurations corresponding to each step of the optimization (case study 1, traditional support fleet, default ice conditions scenario).

| Step | Type 1 | Type 2 | Type 3 | Type 4 | Type 5 | Type 6 | Type 7 | Type 8 | Type 9 | Type 10 | Type 11 | Objective |
|---|---|---|---|---|---|---|---|---|---|---|---|---|
| 1 | 2 | 1 | 2 | 4 | 2 | 0 | 4 | 1 | 5 | 4 | 4 | 123.5 |
| 2 | 4 | 4 | 3 | 3 | 0 | 0 | 1 | 5 | 1 | 3 | 2 | 112.7 |
| 3 | 4 | 4 | 3 | 3 | 0 | 0 | 0 | 2 | 1 | 3 | 2 | 96.9 |
| 4 | 2 | 4 | 3 | 3 | 0 | 0 | 0 | 2 | 1 | 3 | 2 | 89.1 |
| 5 | 2 | 4 | 3 | 3 | 0 | 0 | 0 | 2 | 0 | 3 | 2 | 84.9 |
| 6 | 2 | 4 | 3 | 3 | 0 | 0 | 0 | 1 | 0 | 3 | 2 | 80.8 |
| 7 | 4 | 3 | 1 | 0 | 1 | 0 | 2 | 1 | 2 | 3 | 1 | 78.9 |
| 8 | 4 | 2 | 1 | 0 | 1 | 0 | 2 | 1 | 2 | 3 | 1 | 74.7 |
| 9 | 4 | 2 | 1 | 0 | 1 | 0 | 2 | 1 | 1 | 3 | 1 | 70.7 |
| 10 | 4 | 2 | 1 | 0 | 1 | 0 | 2 | 1 | 1 | 2 | 1 | 67.3 |
| 11 | 4 | 1 | 1 | 0 | 1 | 0 | 2 | 1 | 1 | 2 | 1 | 63.1 |
| 12 | 3 | 0 | 1 | 0 | 1 | 0 | 2 | 1 | 1 | 2 | 1 | 54.6 |
| 13 | 3 | 2 | 2 | 0 | 0 | 0 | 2 | 0 | 1 | 0 | 1 | 52.7 |
| 14 | 2 | 2 | 2 | 0 | 0 | 0 | 2 | 0 | 1 | 0 | 1 | 48.8 |
| 15 | 2 | 1 | 2 | 0 | 0 | 0 | 2 | 0 | 1 | 0 | 1 | 44.4 |
| 16 | 2 | 1 | 2 | 0 | 0 | 0 | 2 | 0 | 0 | 0 | 1 | 40.8 |
| 17 | 2 | 1 | 1 | 0 | 0 | 0 | 2 | 0 | 0 | 0 | 1 | 37.6 |
| 18 | 2 | 1 | 0 | 0 | 0 | 0 | 2 | 0 | 0 | 0 | 1 | 34.8 |
| 19 | 1 | 0 | 0 | 0 | 2 | 0 | 0 | 0 | 0 | 1 | 3 | 33.9 |
| 20 | 1 | 1 | 0 | 0 | 0 | 0 | 1 | 0 | 1 | 1 | 1 | 31.9 |
| 21 | 1 | 1 | 0 | 0 | 0 | 0 | 0 | 0 | 0 | 1 | 2 | 27.1 |
| 22 | 1 | 0 | 0 | 0 | 1 | 0 | 0 | 0 | 0 | 0 | 3 | 25.6 |
| 23 | 1 | 0 | 0 | 0 | 0 | 0 | 0 | 0 | 0 | 0 | 4 | 25.0 |
| 24 | 1 | 0 | 0 | 0 | 0 | 0 | 0 | 0 | 1 | 2 | 1 | 24.5 |
| 25 | 1 | 0 | 0 | 0 | 0 | 0 | 0 | 1 | 0 | 2 | 1 | 24.4 |
| 26 | 1 | 0 | 0 | 0 | 0 | 0 | 0 | 0 | 0 | 2 | 2 | 24.1 |